\documentstyle[preprint,aps,eqsecnum,epsf]{revtex}

\tightenlines

\begin{document}
\title{Glueball Mass Predictions of the Valence Approximation to Lattice QCD}
\author{A.\ Vaccarino and D.\ Weingarten}
\address{IBM Research,
P.O. Box 218, Yorktown Heights, NY 10598}
\maketitle

\begin{abstract}
We evaluate the infinite volume, continuum limit of glueball masses in
the valence (quenched) approximation to lattice QCD. For the lightest
scalar and tensor states we obtain masses of 
$1648 \pm 58$ MeV and $2267 \pm 104$ MeV, respectively.
\end{abstract}


\narrowtext

\section{Introduction}\label{sect:intro}

In recent articles we described calculations of the infinite volume,
continuum limit of scalar and tensor glueball masses in the valence
(quenched) approximation to lattice QCD \cite{masses,review}.  For a
single value of lattice spacing and lattice volume, we reported also a
calculation of the the decay coupling constants of the lightest scalar
glueball to pairs of pseudoscalar mesons. The mass and decay
calculations combined support the identification of $f_0(1710)$ as
primarily composed of the lightest scalar
glueball~\cite{glbdecay,contadv}. Evaluation of the mass of the lightest
scalar quarkonium states and of quarkonium-glueball mixing
amplitudes~\cite{lat98mix} then yield a glueball component for
$f_0(1710)$ of $73.8 \pm 9.5$ \%.  In the present article, we describe
the glueball mass data of Ref. \cite{masses} in greater detail along
with an improved evaluation of the mass predictions which follow from
these data.  For the scalar and tensor glueball masses we obtain $1648
\pm 58$ MeV and $2267 \pm 104$ MeV, respectively.

The valence approximation, on which our results depend, may be viewed as
replacing the momentum dependent color dielectric constant arising from
quark-antiquark vacuum polarization with its zero-momentum limit
\cite{valence} and, for flavor singlet mesons, shutting off transitions
between valence quark-antiquark pairs and gluons.  The valence
approximation is expected to be fairly reliable for low lying flavor
nonsinglet hadron masses, which are determined largely by the low
momentum behavior of the chromoelectric field. This expectation is
supported by recent valence approximation
calculations~\cite{hadrons,tsukuba} of the masses of the lowest flavor
multiplets of spin 1/2 and 3/2 baryons and pseudoscalar and vector
mesons.  The predicted masses are all within about 10\% of
experiment. For the lowest valence approximation glueball masses, the
error arising from the valence approximation's omission of the momentum
dependence of quark-antiquark vacuum polarization we thus also expect to
be 10\% or less.  Refs.~\cite{review,lat96} show this error should tend
to lower valence approximation masses below those of full QCD.  For
flavor singlet configurations whose quantum numbers, if realized as
quarkonium, require nonzero orbital angular momentum, it is shown in
Ref.~\cite{review} that the additional error arising from the valence
approximation's suppression of transitions between valence
quark-antiquark pairs and gluons is likely to introduce an additional
error of the order of 5\% or less.  For the lowest scalar glueball this
error is examined in detail in Ref.~\cite{lat98mix} and found to shift
the valence approximation mass by about 5\% below its value in full QCD.
It is perhaps useful to mention that, for glueball masses, the valence
approximation simply amounts to a reinterpretation of the predictions of
pure gauge theory.

In Section~\ref{sect:ops} we define a family of operators used to
construct glueball propagators.  In Section~\ref{sect:lats} we describe
the set of lattices on which propagators were evaluated and the
algorithms we used to generate gauge configurations and estimate error
bars. In Sections~\ref{sect:scalar} and
\ref{sect:tensor} we present our results for scalar and tensor glueball
propagators, respectively, and masses extracted from these propagators.
In Section~\ref{sect:vol} we estimate the difference between the scalar
and tensor masses we obtain in finite volumes and the corresponding
infinite volume limits. In Section~\ref{sect:cont} we extrapolate scalar
and tensor masses to their continuum limits.  In Section~\ref{sect:comp}
we compare our calculations with work by other
groups~\cite{livertal,Morningstar}. For combined world average valence
approximation scalar and tensor glueball masses we obtain $1656 \pm 47$
MeV and $2302 \pm 62$ MeV, respectively.

\section{Smeared Operators}\label{sect:ops}

We evaluated glueball propagators using operators built out of smeared
link variables. Glueball operators built from link variables with an
optimal choice of smearing couple more weakly to excited glueball states
than do corresponding operators built from unsmeared links.  As a
consequence, the plateau in effective mass plots for optimally smeared
operators begins at a smaller time separation between source and sink
operators, extends over a larger number of time units, and yields a
fitted mass with smaller statistical noise than would be obtained from
operators made from unsmeared link variables.  Examples of the
improvements which we obtained by a choice of smeared operators with be
given in Section~\ref{sect:scalar} and \ref{sect:tensor}.

Initially, we constructed smeared operators from gauge links fixed to
Coulomb gauge. This method gave adequate results for the largest values
of lattice spacing we considered. As the lattice spacing was made
smaller, however, we found that the computer time required to gauge fix
a large enough ensemble of configurations to obtain useful results
became unacceptably large.  We then switched to a gauge invariant
smearing method.  For the lattice sizes used in our extrapolation to the
continuum limit, the gauge invariant mass results had statistical
uncertainties typically a factor of three smaller than our earlier
Coulomb gauge results.  In the remainder of the present article we
discuss only the gauge invariant results. A summary of our Coulomb gauge
mass calculations is given in Ref.~\cite{masses}.

A family of 
gauge invariant smeared operators we construct following the
adaptation in Ref.~\cite{Gupta} of the smearing method of
Ref.~\cite{APE}.  A related method of gauge invariant smearing is
proposed in Refs.~\cite{Tepper}.
For $n > 0$, $\epsilon > 0$, we define iteratively a sequence of
smeared, space-direction link variable $U_i^{n \epsilon}(x)$, with
$U_i^{0 \epsilon}(x)$ given by the unsmeared link variable $U_i(x)$. Let
$u_i^{(n+1) \epsilon}(x)$ be
\begin{eqnarray}
\label{defun}
u_i^{(n+1) \epsilon}(x) & = & U_i^{n \epsilon}(x) + \epsilon \sum_j
U_j^{n \epsilon}(x) U_i^{n \epsilon}(x + \hat{j}) [U_i^{n 
 \epsilon}(x + \hat{i})]^{\dagger} + \nonumber \\ 
 & & \epsilon \sum_j 
[U_j^{n \epsilon}(x - \hat{j})]^{\dagger} U_i^{n \epsilon}(x - \hat{j}) U_i^{n
\epsilon}(x + \hat{i} - \hat{j}),
\end{eqnarray}
where the sum is over the two space directions $j$ orthogonal to
direction $i$.  The projection of $u_i^{(n+1) \epsilon}(x)$ into SU(3)
defines the new smeared link variable $U_i^{(n+1) \epsilon}(x)$.

To find $U_i^{(n+1) \epsilon}(x)$ we maximize over SU(3)
the target function
\begin{eqnarray}
\label{projtarget}
Re Tr \{U_i^{(n+1) \epsilon}(x) [u_i^{(n+1) \epsilon}(x)]^{\dagger} \}.
\end{eqnarray}
The required maximum is constructed by repeatedly applying an algorithm
related to the Cabbibo-Marinari-Okawa Monte Carlo method.  We begin with
$U_i^{(n+1) \epsilon}(x)$ chosen to be 1.  We then multiply $U_i^{(n+1)
\epsilon}(x)$ by a matrix in the SU(2) subgroup acting only on gauge
index values 1 and 2 chosen to maximize the target function over this
subgroup.  This multiplication and maximization step is repeated for the
SU(2) subgroup acting only on index values 2 and 3, then for the
subgroup acting only on index values 1 and 3. The entire three step
process is then repeated five times. Five repetitions we found
sufficient to produce a $U_i^{(n+1) \epsilon}(x)$ satisfactorily close
to the true maximum of the target function in Eq. ~(\ref{projtarget}).
Iteratively maximizing the target function over SU(2) subgroups turns
out to be much easier to program than a direct maximization over all of
SU(3). The additional computer time required for this iterative
maximization, on the other hand, was a negligible fraction of the total
time required for our calculation.

From the $U_i^{n \epsilon}(x)$ we construct $W_{k l}^{n \epsilon s}(x)$
by taking the trace of the product of $U_i^{n \epsilon}(x)$ around the
boundary of an $s \times s$ square of links beginning in the $k$
direction. The sum of $W_{k l}^{n \epsilon s}(x)$ over all sites with a
fixed time value $t$ gives the zero-momentum loop variable $W_{k l}^{n
\epsilon s}(t)$.

For each triple $(n, \epsilon, s)$,
a field coupling the vacuum only to zero-momentum scalars states is
\begin{eqnarray}
\label{defscalar}
S^{n \epsilon s}(t) & = & \sum_{i \ne j} Re W_{i j}^{n \epsilon s}( t),
\end{eqnarray}
where the sums are over space directions $i$ and $j$. 
A possible choice of the two independent operators
coupling the vacuum only to zero-momentum tensor states is
\begin{eqnarray}
\label{deftensor}
T_1^{n \epsilon s}(t) & = & 2 Re W_{12}^{n \epsilon s}( t) - 
Re W_{23}^{n \epsilon s}( t) - Re W_{31}^{n \epsilon s}( t) \nonumber \\ 
T_2^{n \epsilon s}(t) & = & \sqrt{3} Re W_{23}^{n \epsilon s}( t) - 
\sqrt{3} Re W_{31}^{n \epsilon s}( t).  
\end{eqnarray}
The optimal choice of $(n, \epsilon, s)$ for each operator and lattice
spacing will be considered in the next section.

\section{Lattices, Monte Carlo Algorithm and Error Bars}\label{sect:lats}

The set of lattices on which we evaluated scalar and tensor glueball
propagators is listed in Table~\ref{tab:lats}.  

On each lattice, an ensembles of gauge configurations was generated by a
combination of the Cabbibo-Marinari-Okawa algorithm and the overrelaxed
method of Ref.~\cite{Brown}. To update a gauge link we first performed a
microcanonical update in the SU(2) subgroup acting on gauge indices 1
and 2. This was then repeated for the SU(2) subgroup acting on indices 2
and 3, and the subgroup acting on indices 1 and 3. These three update
steps were then repeated on each link of the lattice. After four lattice
sweeps each consisting of the three microcanonical steps on each link,
we carried out one Cabibbo-Marinari-Okawa sweep of the full lattice.

At least 10,000 sweeps were used in each case to generate an initial
equilibrium configuration. The number of sweeps skipped between each
configuration used to calculate propagators and the total number of
configurations in each ensemble are listed in the third and fourth
columns, respectively, of Table~\ref{tab:lats}. Although the number of
sweeps skipped in each case was not sufficient to permit successive
configurations to be treated as statistically independent, we found
successive configurations to be sufficiently independent to justify the
cost of evaluating glueball operators.

For the propagators, effective masses and fitted masses to be discussed
in Sections~\ref{sect:scalar} and \ref{sect:tensor}, we determined
statistical uncertainties by the bootstrap method \cite{Efron}. The
bootstrap algorithm can be applied directly, however, only to determine
the uncertainties in quantities obtained from an ensemble whose
individual members are statistically independent.  We therefore
partitioned each ensemble of correlated gauge configurations into
successive disjoint bins with a fixed bin size.  Bootstrap ensembles
were then formed by randomly choosing a number of entire bins equal to
the number of bins in the original partitioned ensemble.  For bins
sufficiently large, propagator averages found on distinct bins will be
nearly independent.  It follows that for large enough bins, the binned
bootstrap estimate of errors will be reliable.  It is not hard to show
that once bins are made large enough to produce nearly independent bin
averages, further increases in bin size will leave bootstrap error
estimates nearly unchanged. The only variation in errors as the bin size
is increased further will come from statistical fluctuations in the
error estimates themselves.  To determine the required bin size for a
particular error estimate to be reliable we applied the bootstrap method
repeatedly with progressively larger bin sizes until the estimated error
became nearly independent of bin size. The final bin size we adopted for
each lattice, chosen to be large enough for all of the error estimates
done on that lattice, is given in fifth column of Table~\ref{tab:lats}.

\section{Scalar Propagators and Masses}\label{sect:scalar}

From the scalar operator of Eq.~\ref{defscalar}, a propagator for scalars
is defined to be 
\begin{eqnarray}
\label{defPS}
P^{n \epsilon s}_S( t_1 - t_2 )  =  
< S^{n \epsilon s}( t_1) S^{n \epsilon s}( t_2) > 
- < S^{n \epsilon s}( t_1)> <S^{n \epsilon s}( t_2) >.
\end{eqnarray}
To reduce statistical noise, $P^{n \epsilon s}_S( t_1 - t_2 )$ is then
averaged over reflections and time direction displacements of $t_1$ and
$t_2$.

The collection of values of smearing iterations n, smearing parameter
$\epsilon$, and loop size $s$ for which propagators were evaluated for
each lattice are given in Table~\ref{tab:nepsilons}.  At $\beta$ of 5.70,
and at $\beta$ of 5.93 on the lattice $16^3 \times 24$, we ran with
relatively larger ranges of parameters to try to find values which
coupled efficiently to the lightest scalar glueball. For other lattices,
the parameter range was then narrowed to choices which, in physical
units, were about the same as the value range which gave best results at
$\beta$ of 5.93.

From the existence of a self-adjoint, positive, bounded transfer matrix
for lattice QCD, it follows that a spectral resolution can be constructed
for  $P^{n \epsilon s}_S( t )$,
\begin{eqnarray}
\label{scalarspec}
P^{n \epsilon s}_S( t ) & = & \sum_i Z_i \{ exp( -E_i t) + exp[ -E_i (L -
t)] \}, \nonumber \\
Z_i & = & | < i | S^{n \epsilon s}( 0) | vacuum >|^2,
\end{eqnarray}
where the sum is over all zero-momentum, scalar states $< i |$, $E_i$ is
the energy of $< i |$, and $L$ is the lattice period in the time
direction.  For large values of $t$ and $L$, the sum in
Eq.~(\ref{scalarspec}) approaches the asymptotic form
\begin{eqnarray}
\label{scalarasym}
P^{n \epsilon s}_S( t )  \rightarrow Z  \{ exp( -m t) + exp[ -m (L -
t)] \} 
\end{eqnarray}
where $m$ is the smallest $E_i$ and thus the mass of the lightest scalar
glueball and $Z$ is the corresponding $Z_i$. Fitting $P^{n \epsilon s}_S(
t )$ to the asymptotic form in Eq.~(\ref{scalarasym}) at $t$ and $t+1$
gives the scalar effective mass $m( t)$, which at large $t$ approaches
$m$.

To extract values of $m$ from our data sets, we began by examining
effective mass graphs to find combinations of $n$ and $s$ for which
$m(t)$ shows a plateau at $t$ values for which we have data, and to
determine which of these combination of $n$ and $s$ has the best
plateaus.

\begin{figure}
\epsfxsize=\textwidth
\epsfbox{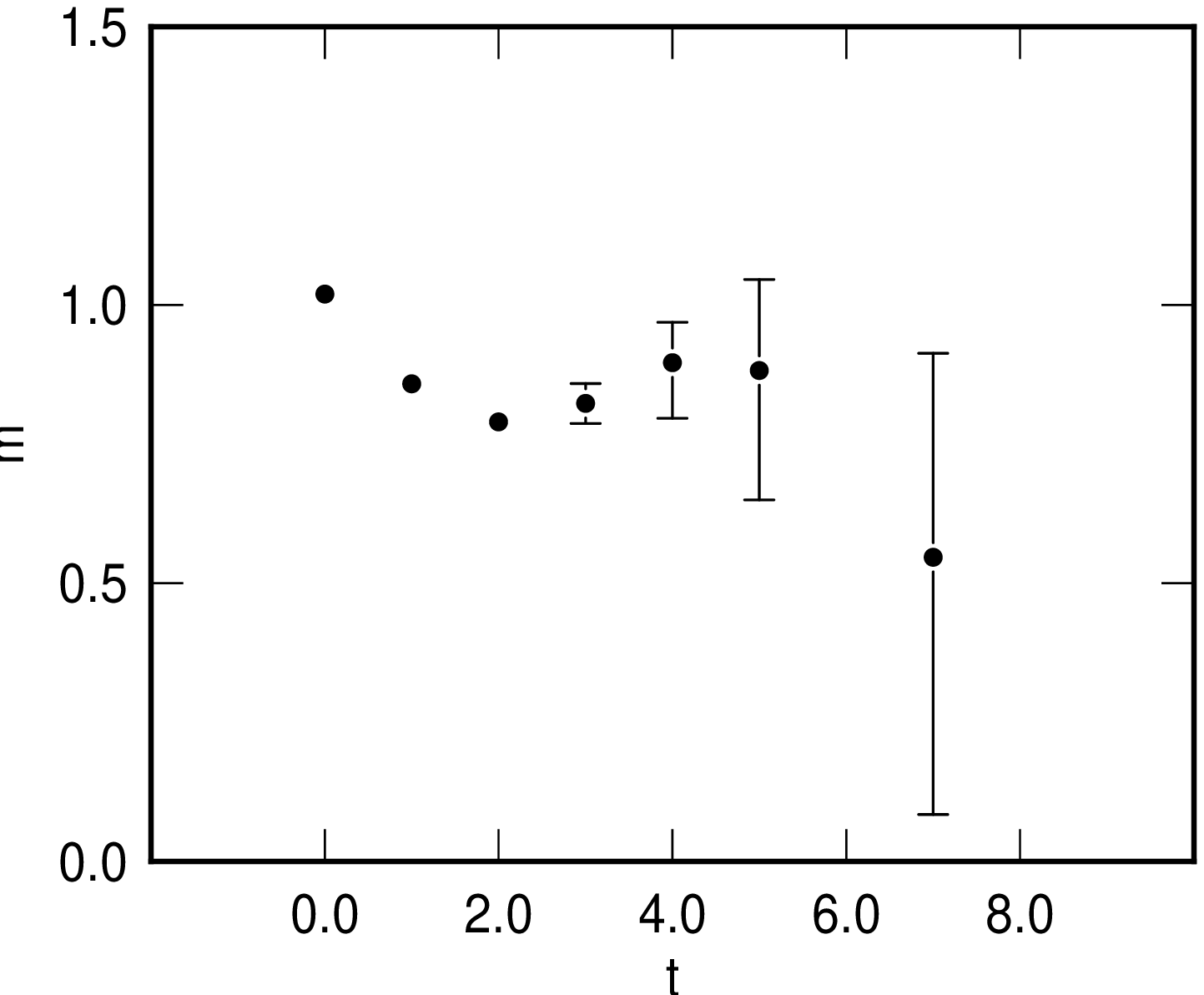}
\caption{Scalar effective mass as a function of $t$ for the lattice
$16^3 \times 24$ at $\beta$ of 5.93 using the smeared operator with
$n$ of 5, $\epsilon$ of 1.0 and $s$ of 3.}
\label{fig:zx16b593n5s3}
\end{figure}

\begin{figure}
\epsfxsize=\textwidth
\epsfbox{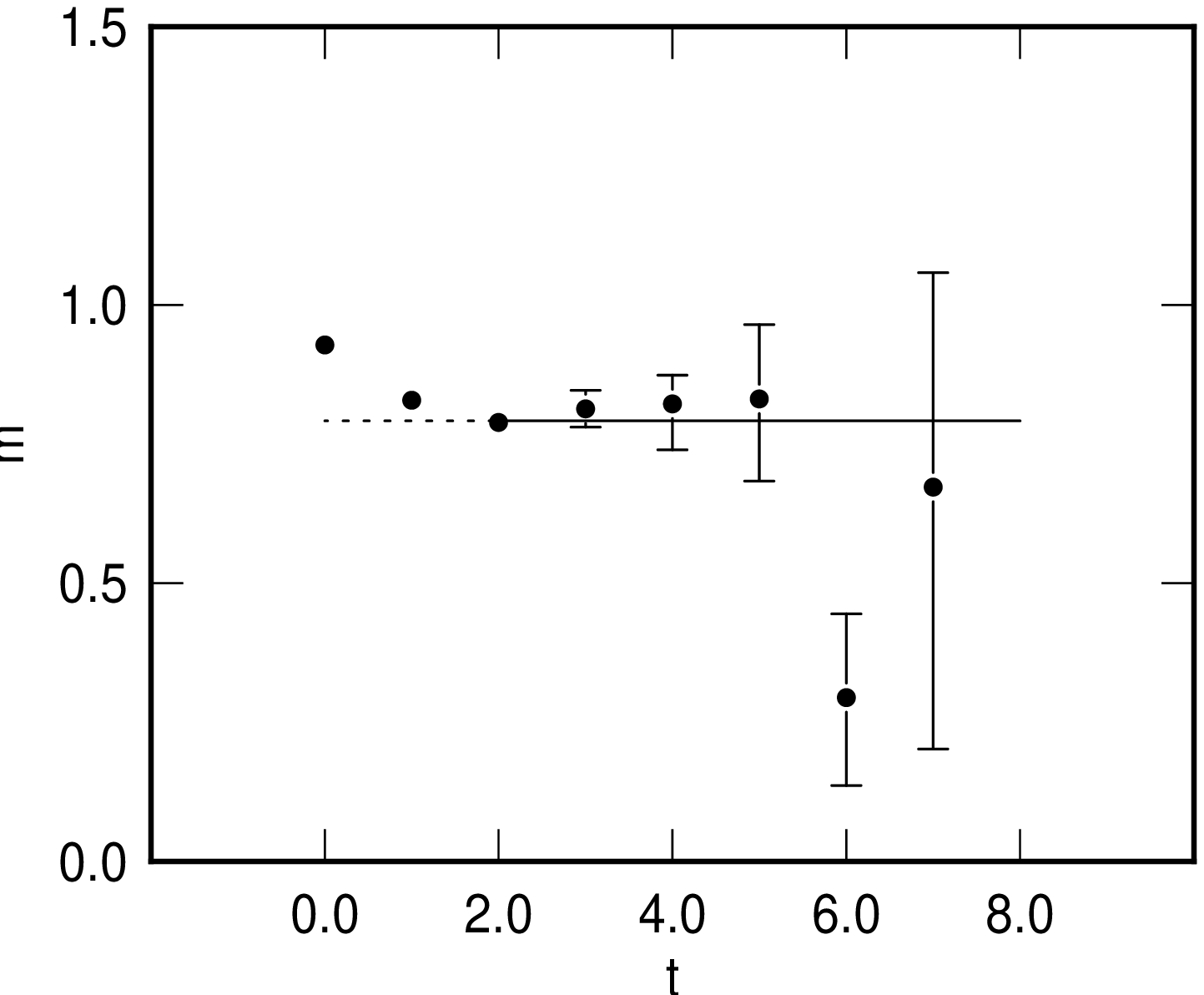}
\caption{Scalar mass fit and
scalar effective mass as a function of $t$ for the
lattice $16^3 \times 24$ at $\beta$ of 5.93 using the smeared operator
with $n$ of 5, $\epsilon$ of 1.0 and $s$ of 4.}
\label{fig:zx16b593n5s4}
\end{figure}

\begin{figure}
\epsfxsize=\textwidth
\epsfbox{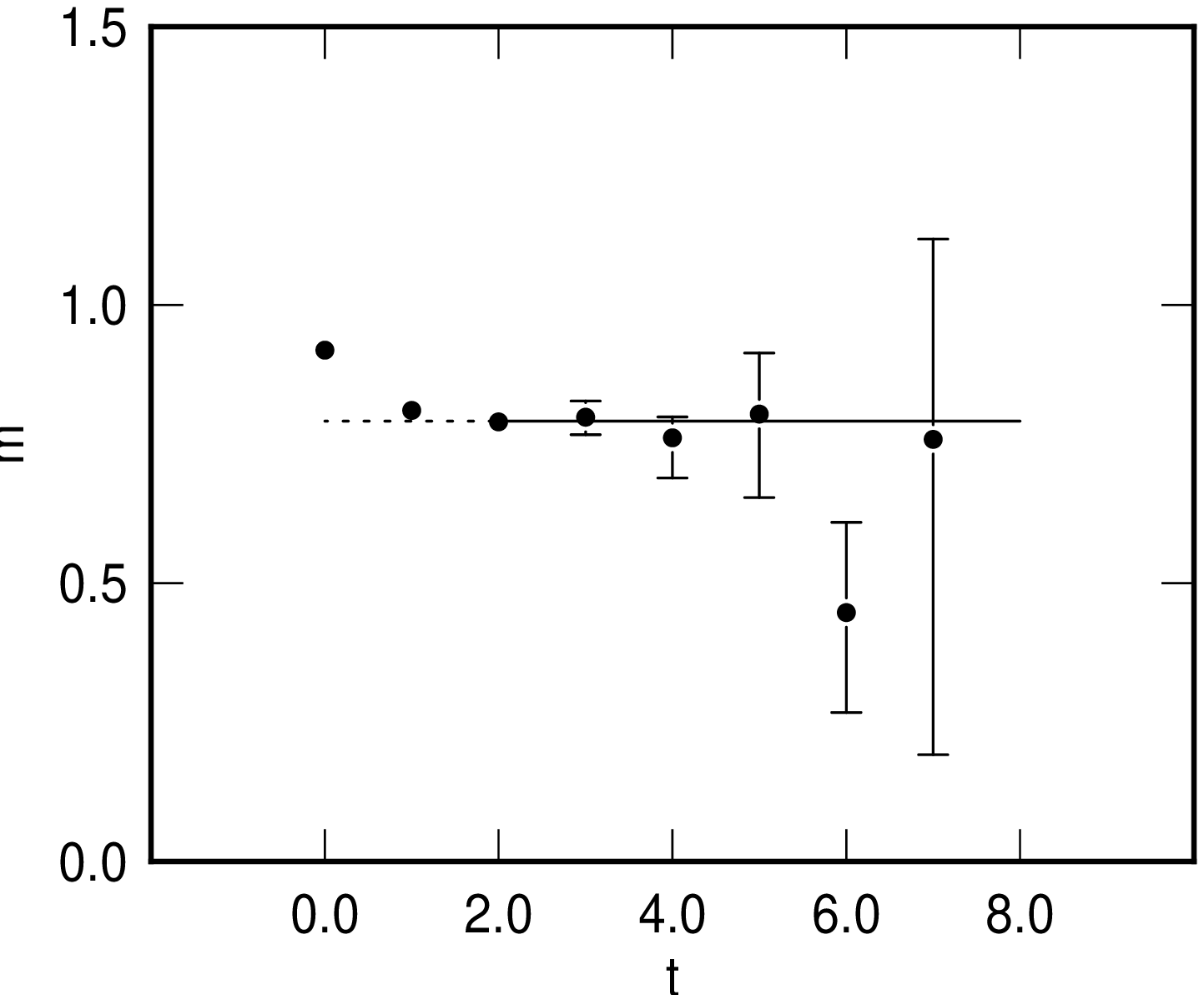}
\caption{Scalar mass fit and
scalar effective mass and fitted mass as a function of $t$ for the lattice
$16^3 \times 24$ at $\beta$ of 5.93 using the smeared operator with
$n$ of 5, $\epsilon$ of 1.0 and $s$ of 5.}
\label{fig:zx16b593n5s5}
\end{figure}

\begin{figure}
\epsfxsize=\textwidth
\epsfbox{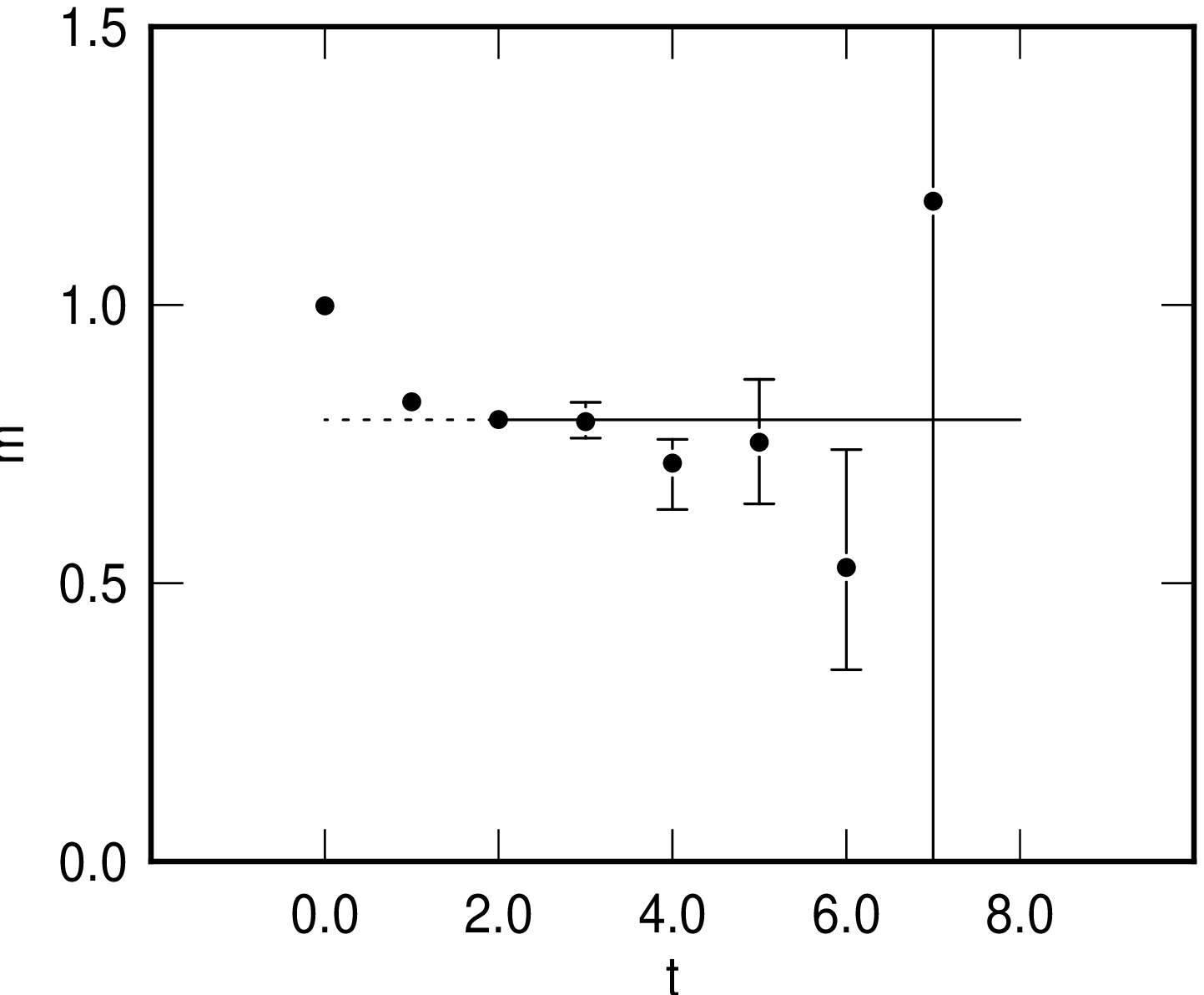}
\caption{Scalar mass fit and
scalar effective mass as a function of $t$ for the lattice
$16^3 \times 24$ at $\beta$ of 5.93 using the smeared operator with
$n$ of 5, $\epsilon$ of 1.0 and $s$ of 6.}
\label{fig:zx16b593n5s6}
\end{figure}

\begin{figure}
\epsfxsize=\textwidth
\epsfbox{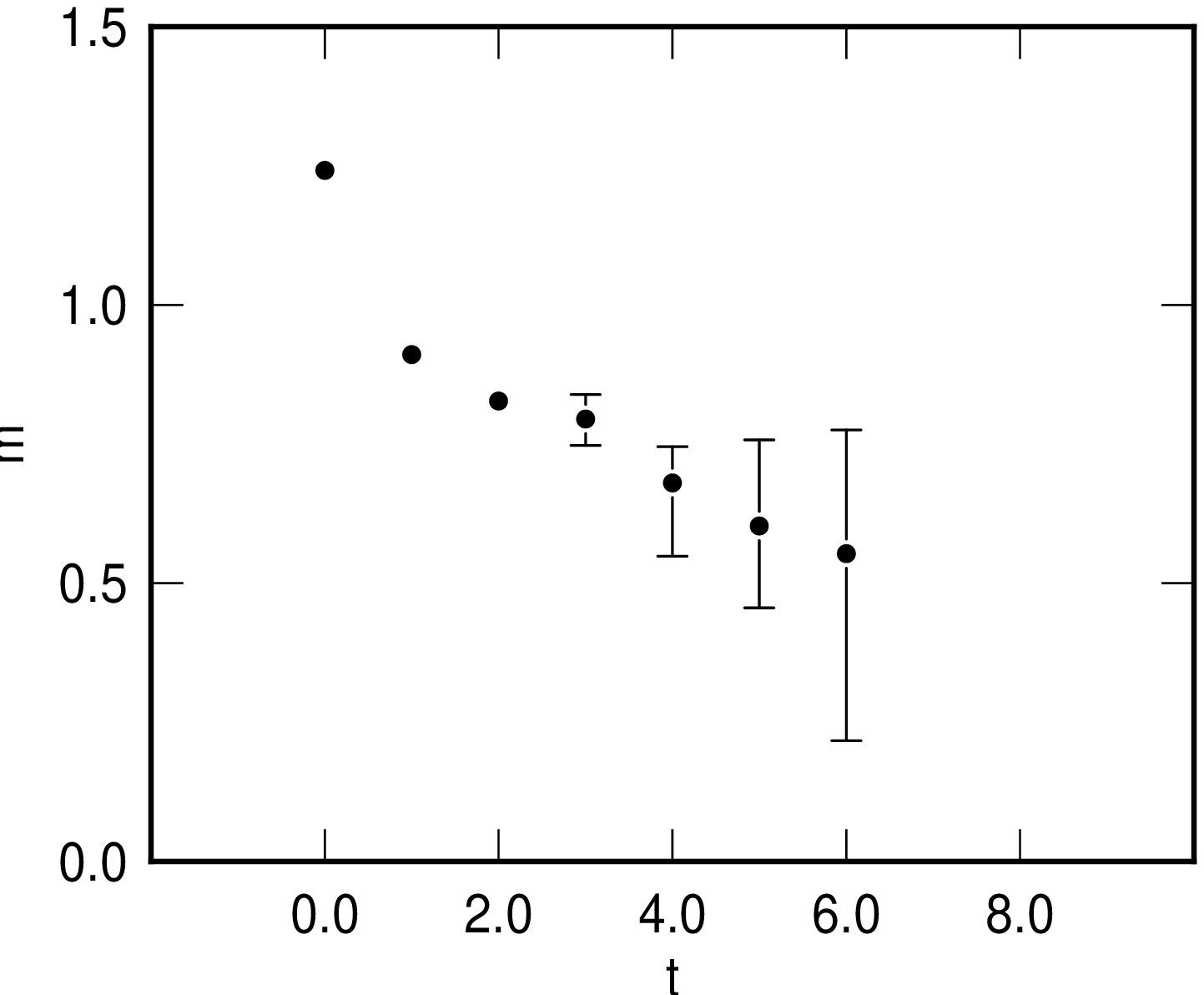}
\caption{Scalar effective mass as a function of $t$ for the lattice
$16^3 \times 24$ at $\beta$ of 5.93 using the smeared operator with
$n$ of 5, $\epsilon$ of 1.0 and $s$ of 7.}
\label{fig:zx16b593n5s7}
\end{figure}

\begin{figure}
\epsfxsize=\textwidth
\epsfbox{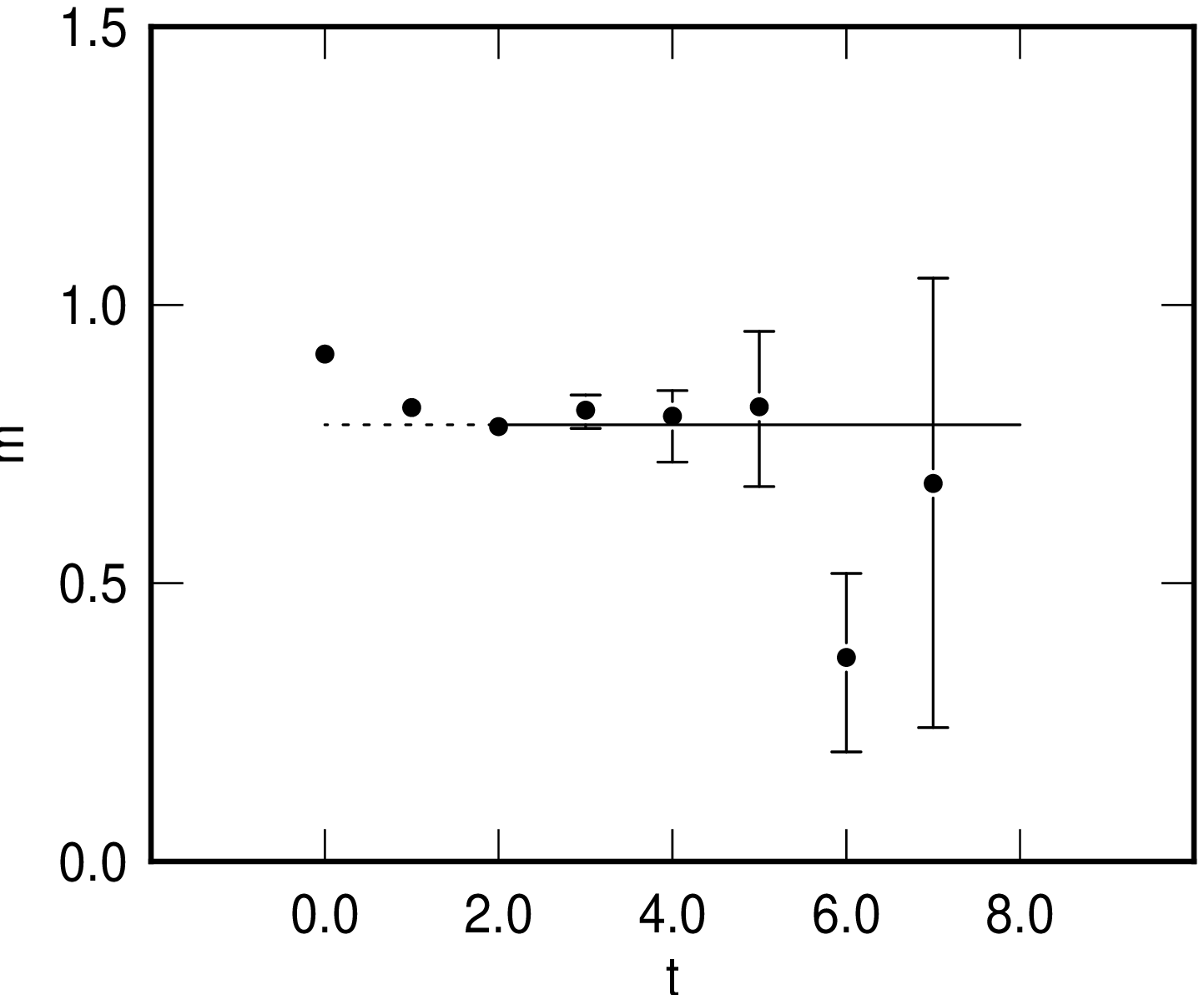}
\caption{Scalar mass fit and
scalar effective mass as a function of $t$ for the
lattice $16^3 \times 24$ at $\beta$ of 5.93 using the smeared operator
with $n$ of 6, $\epsilon$ of 1.0 and $s$ of 4.}
\label{fig:zx16b593n6s4}
\end{figure}

\begin{figure}
\epsfxsize=\textwidth
\epsfbox{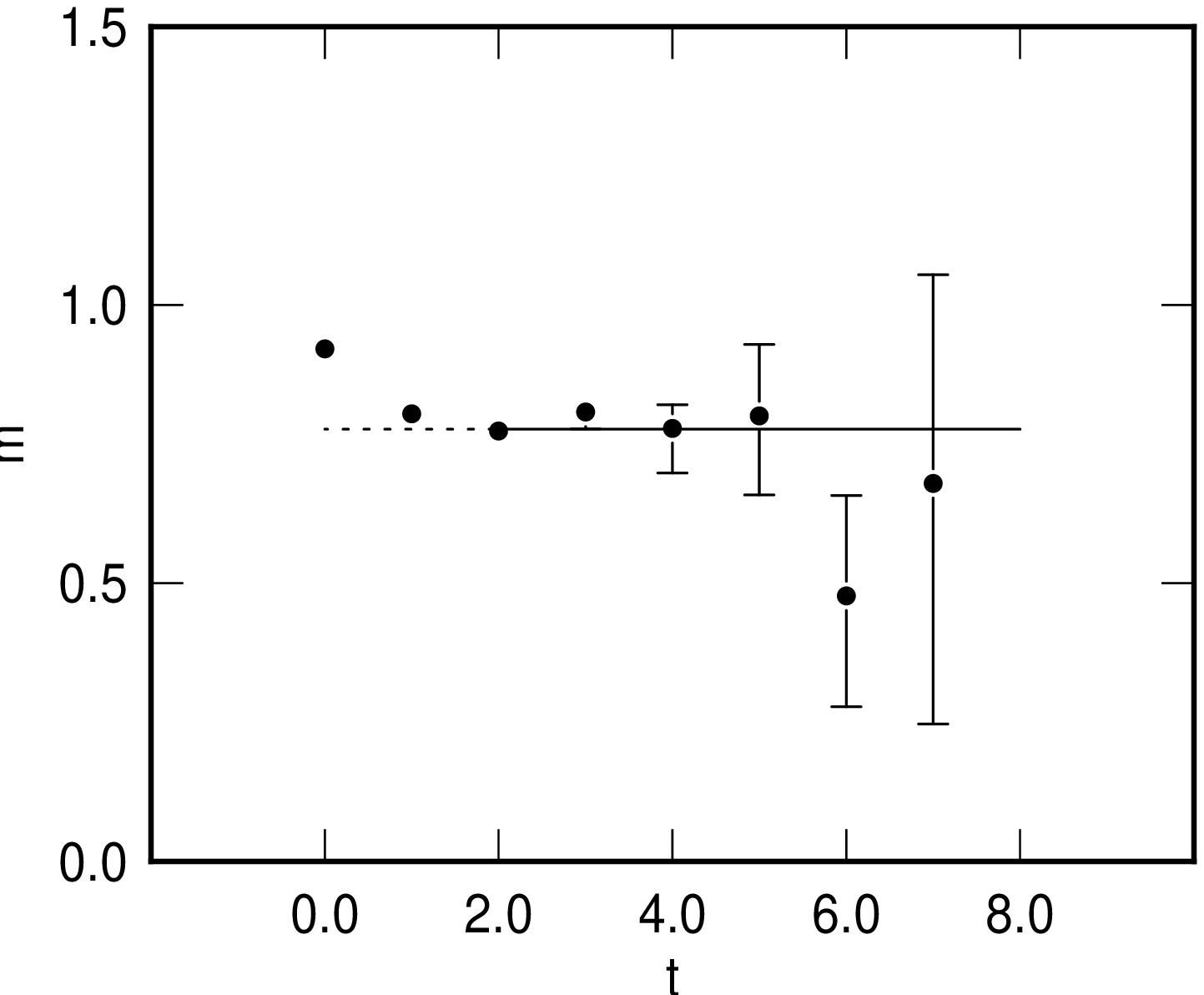}
\caption{Scalar mass fit and
scalar effective mass as a function of $t$ for the
lattice $16^3 \times 24$ at $\beta$ of 5.93 using the smeared operator
with $n$ of 7, $\epsilon$ of 1.0 and $s$ of 4.}
\label{fig:zx16b593n7s4}
\end{figure}

\begin{figure}
\epsfxsize=\textwidth
\epsfbox{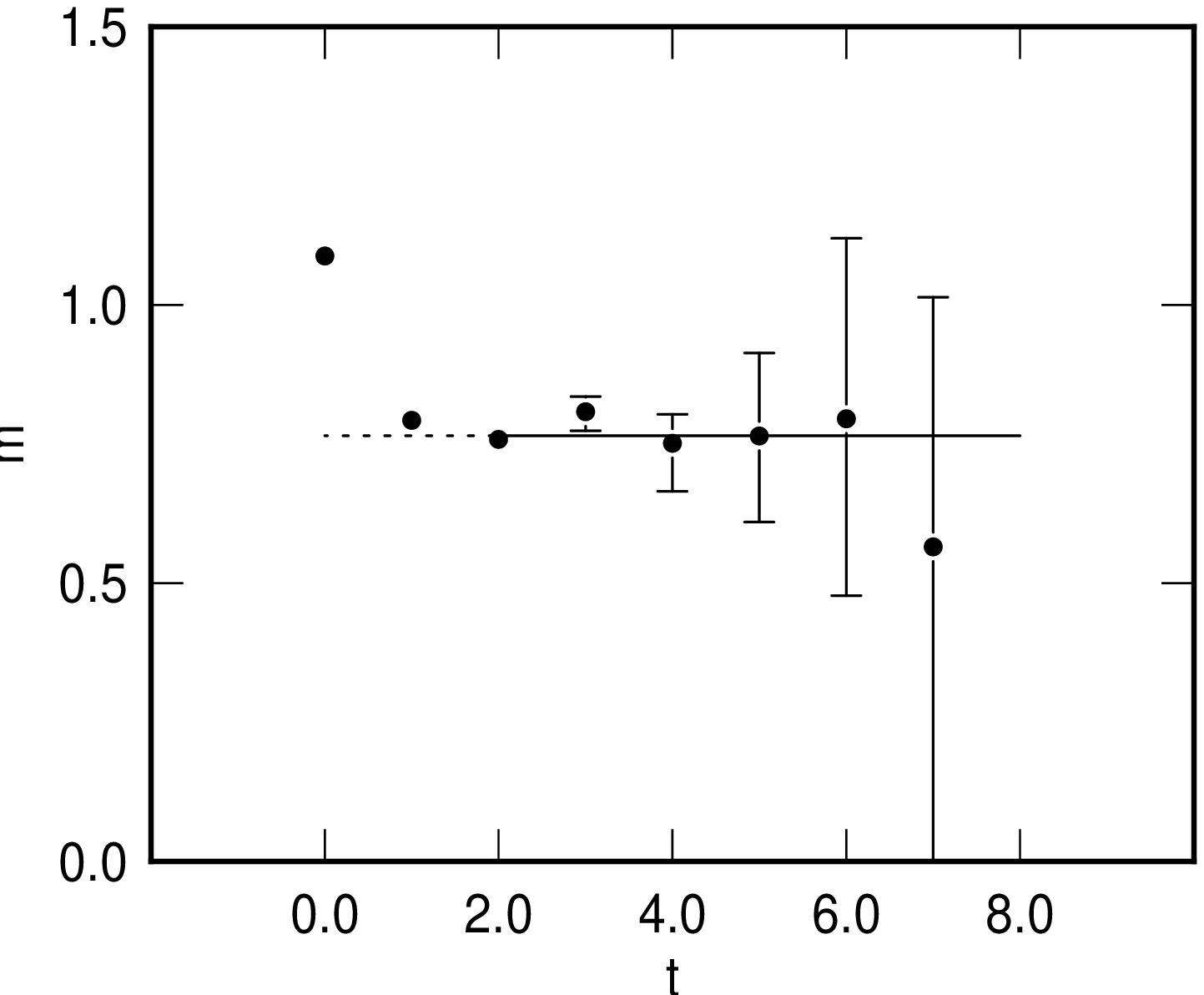}
\caption{Scalar mass fit and
scalar effective mass as a function of $t$ for the
lattice $16^3 \times 24$ at $\beta$ of 5.93 using the smeared operator
with $n$ of 8, $\epsilon$ of 1.0 and $s$ of 4.}
\label{fig:zx16b593n8s4}
\end{figure}

Among the data sets used in our final extrapolation of the scalar mass
to zero lattice spacing, we included the largest range of values of $n$ and
$s$ for the lattice $16^3 \times 24$ with $\beta$ of 5.93.  Scalar
effective masses obtained for this case with $n$ of $5$ and $s$ of $3 -
7$ are shown in Figures~\ref{fig:zx16b593n5s3} -
\ref{fig:zx16b593n5s7}, respectively.  As the loop size $s$ is
increased, initially the effective mass graphs become flatter, as
shown, for example, by a decrease in the difference between $m(0)$ and $m( 2)$.
It follows that the relative coupling of the corresponding
operators to the lightest scalar glueball increases with $s$.
Beyond $s$ of 5, however, this trend reverse. Thus, as might be 
expected, the relative coupling to the lightest state becomes weaker again
when the loop is made too large.  For $s$ of 7 the effective mass graph
shows no sign of becoming flat even at the largest $t$ for which we have
statistically significant data. For $n$ of 5, the best coupling to the
lightest state appears to occur with $s$ of 4 or 5. For $s$ fixed at 4,
Figure~\ref{fig:zx16b593n5s4} and Figures~\ref{fig:zx16b593n6s4} -
\ref{fig:zx16b593n8s4} show the variation in the effective mass graph as
$n$ runs from 5 to 8, respectively. The difference between $m(0)$ and $m(2)$
is least at $n$ of 6 and then grows again as $n$ is raised toward 8.

In Figures~\ref{fig:zx16b593n5s3} - \ref{fig:zx16b593n8s4} the
statistical uncertainty in effective masses grows as $t$ is made larger
and tends to grow also if $n$ or $s$ is increased. Both of these
phenomena are explained by the discussion in Ref.~\cite{vanBaal} of the
statistical uncertainty in propagators.

\begin{figure}
\epsfxsize=\textwidth
\epsfbox{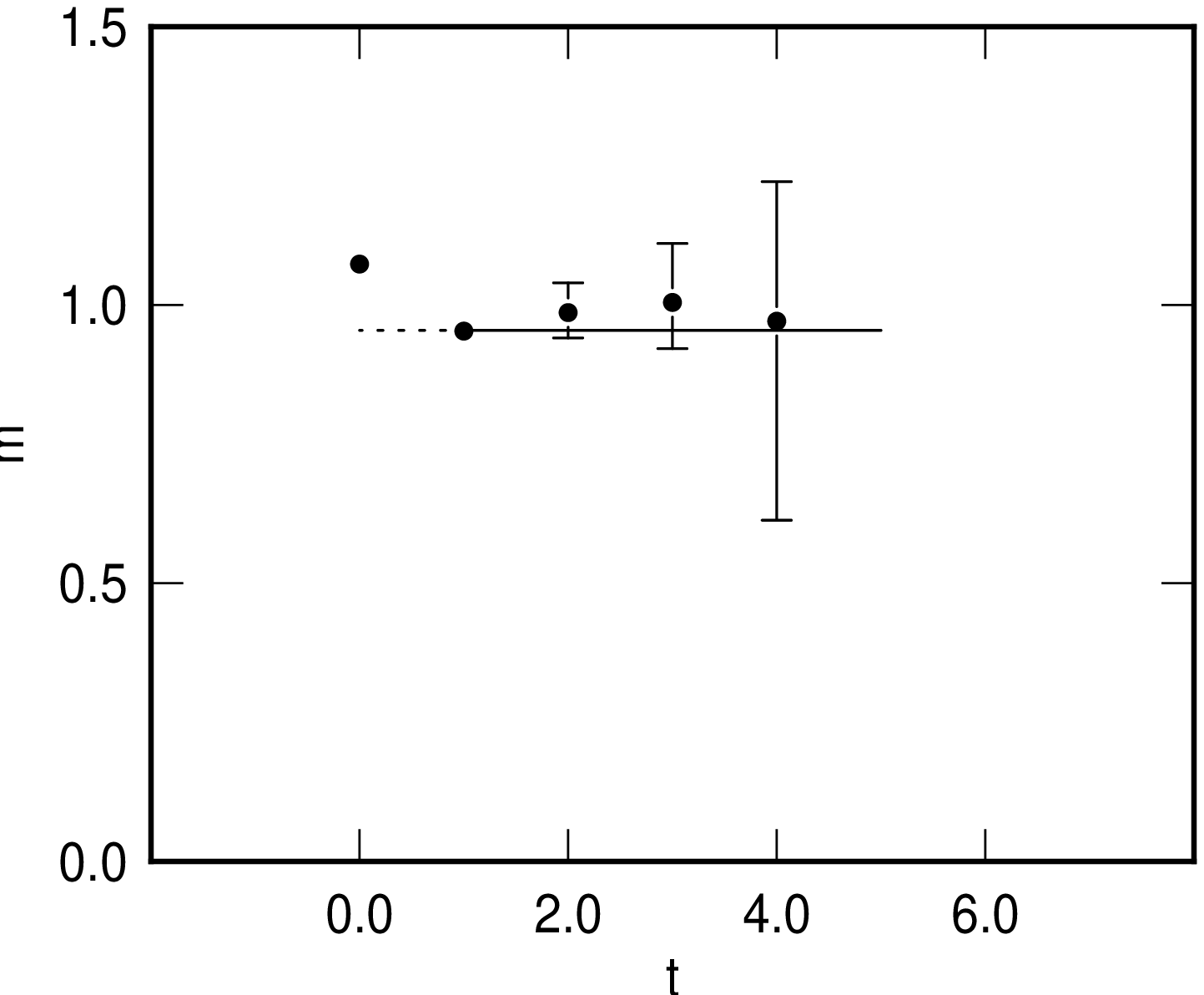}
\caption{Scalar mass fit and
scalar effective mass as a function of $t$ for the
lattice $16^3 \times 24$ at $\beta$ of 5.70 using the smeared operator
with $n$ of 6, $\epsilon$ of 0.25 and $s$ of 2.}
\label{fig:zx16b570n6s2}
\end{figure}

\begin{figure}
\epsfxsize=\textwidth
\epsfbox{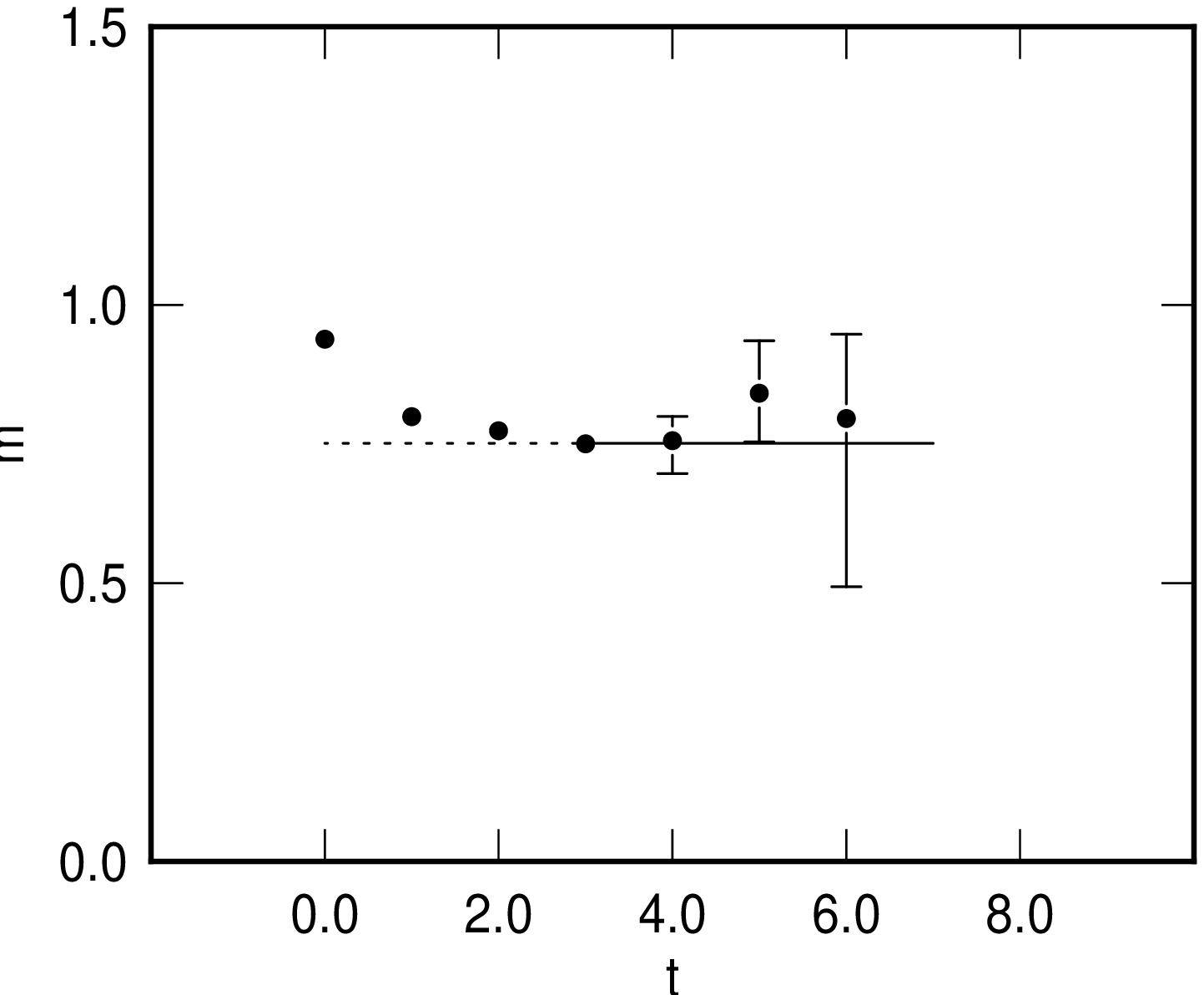}
\caption{Scalar mass fit and
scalar effective mass as a function of $t$ for the
lattice $12^3 \times 24$ at $\beta$ of 5.93 using the smeared operator
with $n$ of 6, $\epsilon$ of 1.0 and $s$ of 6.}
\label{fig:zx12b593n6s6}
\end{figure}

\begin{figure}
\epsfxsize=\textwidth
\epsfbox{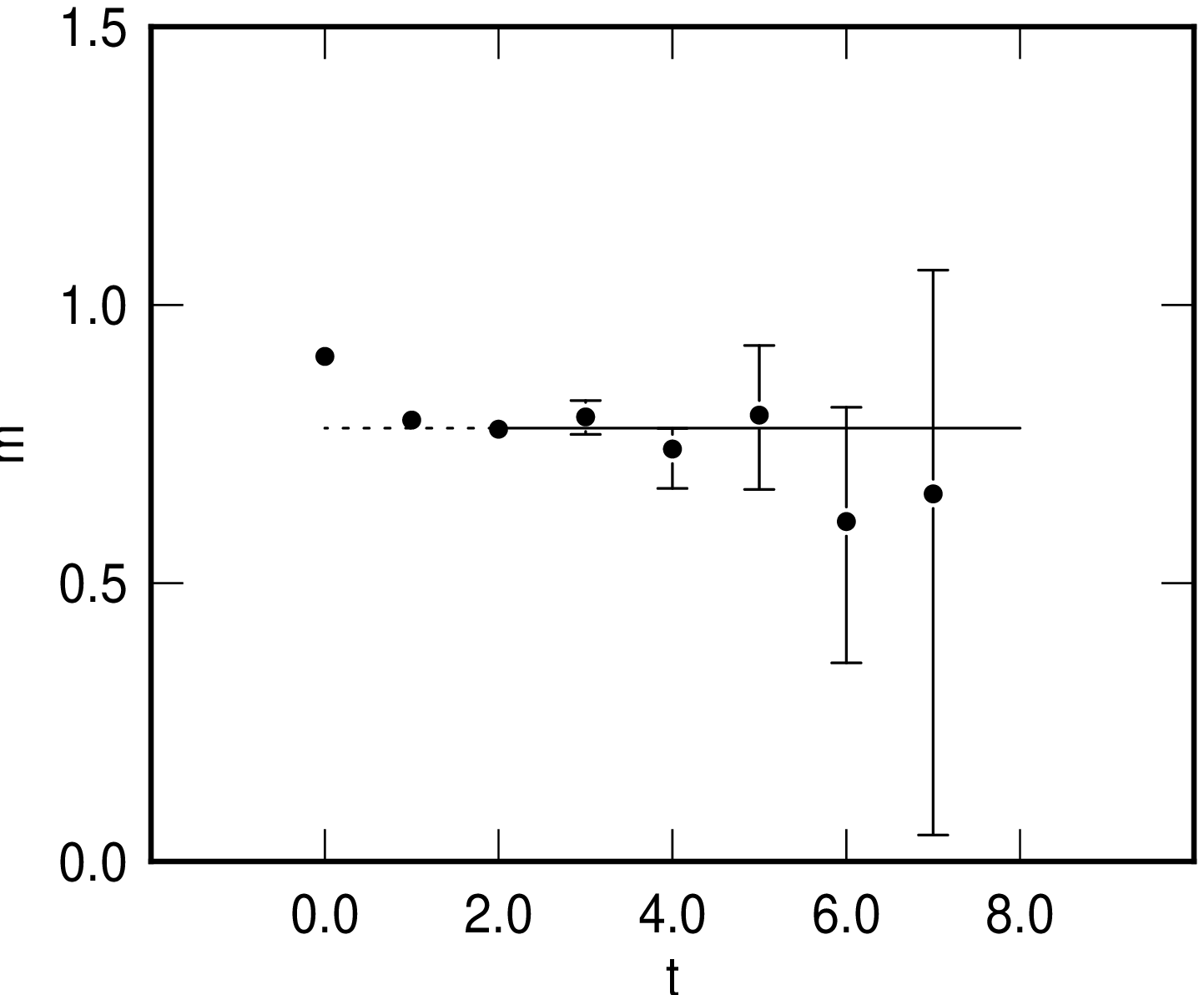}
\caption{Scalar mass fit and
scalar effective mass as a function of $t$ for the
lattice $16^3 \times 24$ at $\beta$ of 5.93 using the smeared operator
with $n$ of 7, $\epsilon$ of 1.0 and $s$ of 5.}
\label{fig:zx16b593n7s5}
\end{figure}

\begin{figure}
\epsfxsize=\textwidth
\epsfbox{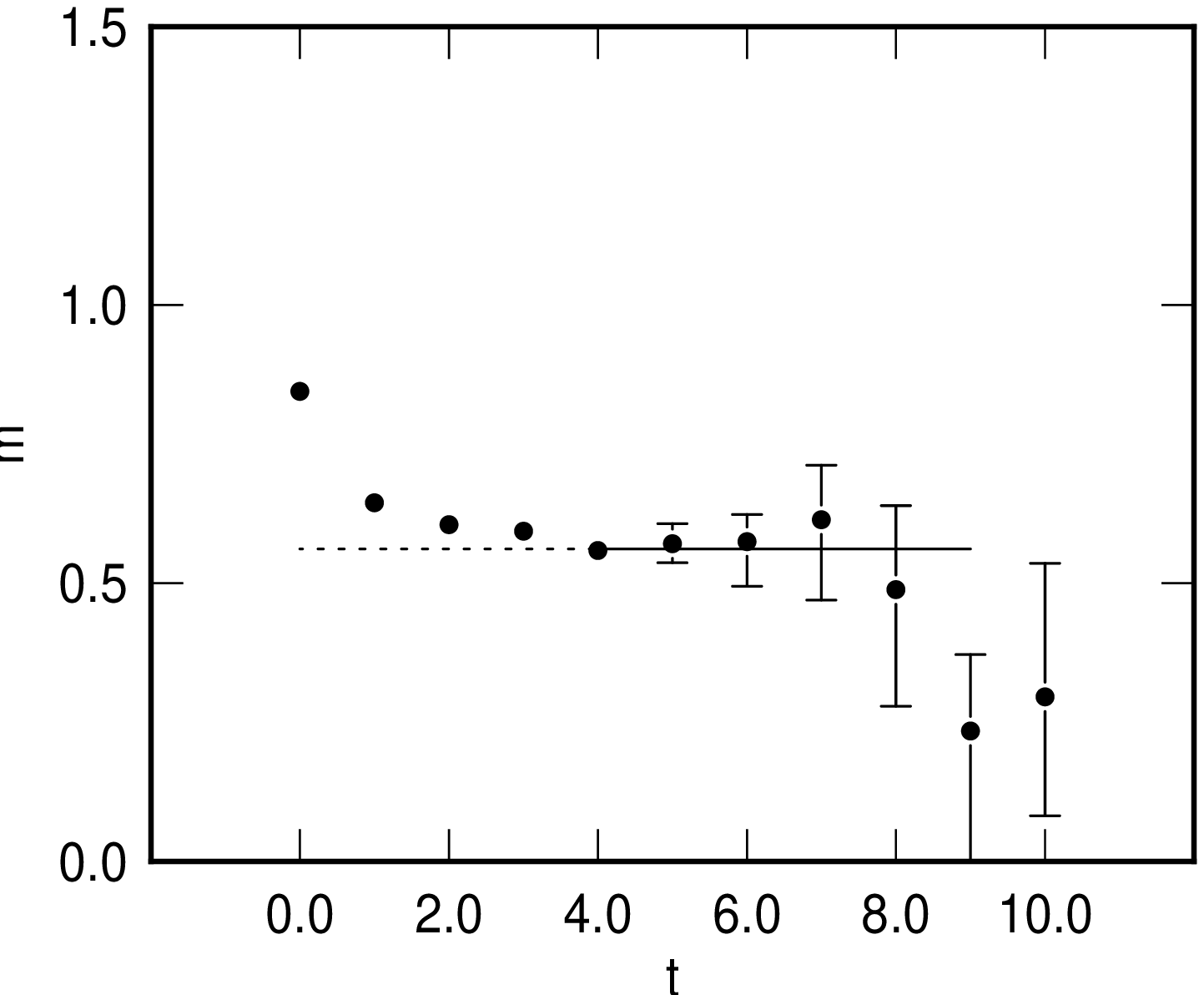}
\caption{Scalar mass fit and
scalar effective mass as a function of $t$ for the
lattice $24^3 \times 36$ at $\beta$ of 6.17 using the smeared operator
with $n$ of 7, $\epsilon$ of 1.0 and $s$ of 8.}
\label{fig:zx24b617n7s8}
\end{figure}

\begin{figure}
\epsfxsize=\textwidth
\epsfbox{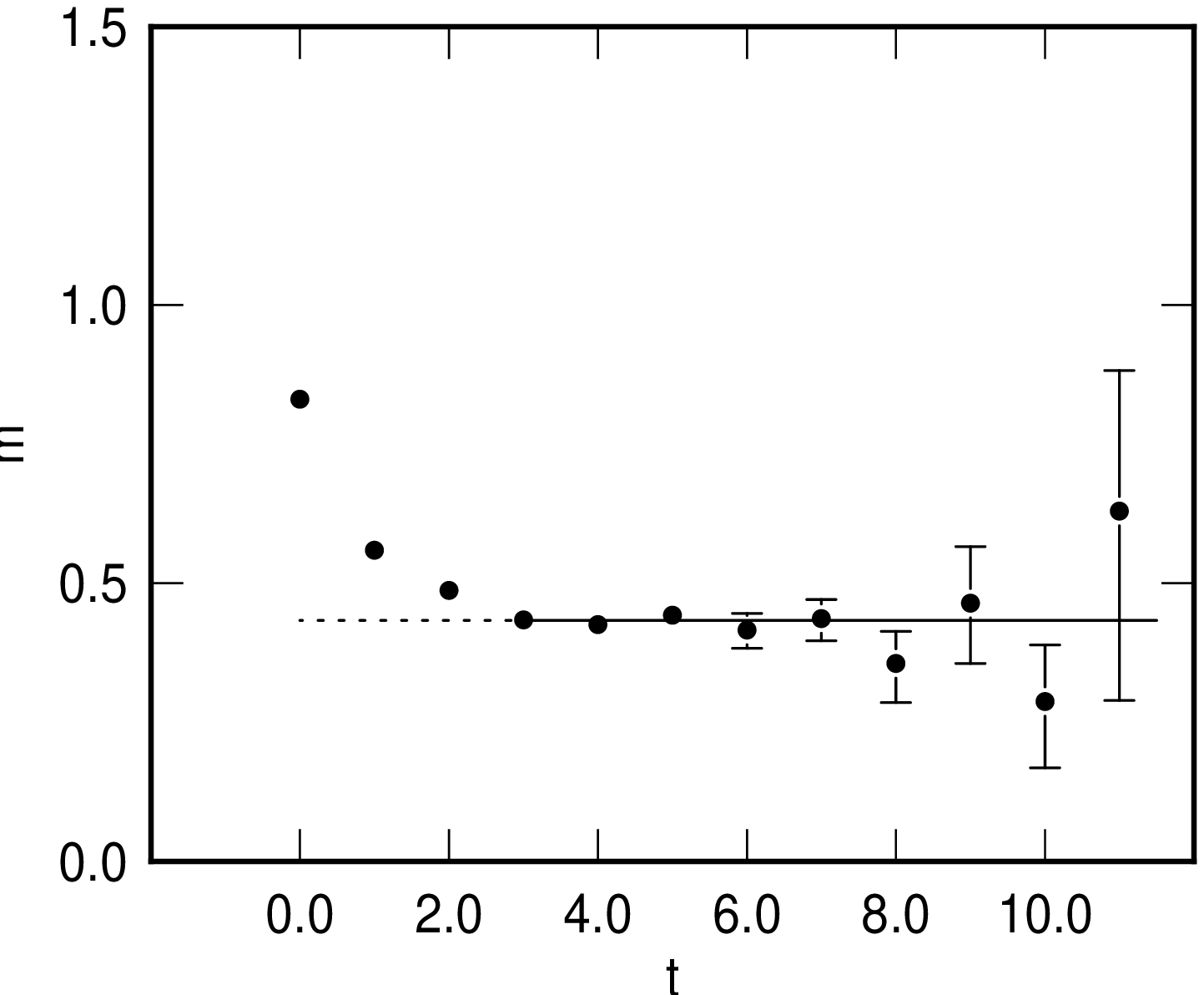}
\caption{Scalar mass fit and
scalar effective mass as a function of $t$ for the
lattice $32^2 \times 30 \times 40$ at $\beta$ of 6.40 using the smeared operator
with $n$ of 8, $\epsilon$ of 1.0 and $s$ of 10.}
\label{fig:zx32b640n8s10}
\end{figure}

Figures~\ref{fig:zx16b570n6s2} - \ref{fig:zx32b640n8s10} show scalar
effective masses for the each of the values of lattice size and $\beta$
listed in Table~\ref{tab:lats}.  The parameters $n$ and $s$ for the data
in Figures~\ref{fig:zx16b570n6s2} - \ref{fig:zx32b640n8s10} are chosen,
for each lattice and $\beta$, from among the set which couple best to
the lightest scalar.

For each combination of lattice size and $\beta$, we determined a final
value of the scalar mass from the collection of propagators for which
the effective mass graph showed at least some evidence of a plateau at
large $t$.  For several different choices of $t$ interval, each
of these propagators was fitted to the asymptotic form in
Eq.~(\ref{scalarasym}) by minimizing the fit's correlated $\chi^2$. The
upper limit of each fitting interval $t_{max}$ we fixed at the largest
$t$ for which we had statistically significant propagator data.  The
lower limit of the fitting interval $t_{min}$ was then progressively
increased from 1 to $t_{max} - 2$. As $t_{min}$ was increased, the
fitted mass and the fit's $\chi^2$ per degree of freedom both generally
decreased and the statistical error bar increased. For each $n$ and $s$,
the final choice of $t_{min}$ we took to be the smallest value for which
the corresponding mass was within the error bars of all the fits with
the same $n$ and $s$ and larger $t_{min}$. Our intent in this procedure
was to extract a mass from the largest time interval for which the
propagator for each combination of $n$ and $s$ was consistent with the
asymptotic form of Eq.~(\ref{scalarasym}).

The solid horizontal lines in Figures~\ref{fig:zx16b593n5s4} -
\ref{fig:zx16b593n5s7} and Figures~\ref{fig:zx16b593n6s4} -
\ref{fig:zx32b640n8s10} show the best fitted mass in each case and
extend over the interval of $t$ on which these fits were made. The
dashed lines in these figures extend the solid lines to smaller $t$ to
show the approach, with increasing $t$, of effective masses to the final
mass values.

For the lattice $16^3 \times 24$ with $\beta$ of 5.93,
Tables~\ref{tab:593zd1} - \ref{tab:593zd4} show the results
of our fits for all the combination of $n$, $s$, $t_{min}$ and $t_{max}$
which we examined.  The best choice of $t_{min}$ and $t_{max}$ turned
out to be 2 and 8, respectively, for all $n$ and $s$.
Tables~\ref{tab:570z} - \ref{tab:640z} show the fitted masses
found with the best choice of $t_{min}$ and $t_{max}$ for all the
lattice sizes, $\beta$, $n$ and $s$ for which our effective
mass data showed a plateau at large $t$.

As expected, for each lattice size and $\beta$, the fitted masses in
Tables~\ref{tab:570z} - \ref{tab:640z} vary with $n$ and $s$ by an
amount generally less than the statistical uncertainty in each mass.
There is also a weak tendency for masses to fall initially with
increasing $n$ and $s$, as the corresponding operator's relative coupling
to the lightest glueball increases. Then, in some cases, when $n$ and
$s$ become too large the coupling to the lightest state decreases, the
fitted masses show some tendency to rise again.  To reduce this small
remaining statistical uncertainty and systematic bias, our final value
of mass for each lattice size and $\beta$ was obtained by an additional
fit of a single common mass to a set of masses from a range of several
$n$ and $s$.  The common mass was chosen to minimize the correlated
$\chi^2$ of the fit of the common mass to the collection of different
best mass values. The correlation matrix among the best mass values was
determined by the bootstrap method.  The set of $n$ and $s$ used in each
final fit was chosen by examining a decreasing sequence of sets,
starting with all $n$ and $s$, and progressively eliminating the smallest
and largest $n$ and $s$ until a $\chi^2$ per degree of freedom below 2.0
was obtained.  The final fit was taken from the largest set of channels
yielding a $\chi^2$ below 2.0. If several sets of equal size gave
$\chi^2$ per degree of freedom below 2.0, we chose among these the set
with smallest $\chi^2$ per degree of freedom.
Tables~\ref{tab:570zf}-~\ref{tab:640zf} show these combined fits and the
set of $n$ and $s$ chosen for the final mass value for each lattice and
$\beta$. In all of these tables, it is clear that once enough of the
largest and smallest $n$ and $s$ are eliminated to give an acceptable
$\chi^2$ per degree of freedom, the fitted values vary only by small
fractions of their statistical uncertainty as additional changes are
made in the set of $n$ and $s$. The final mass values are collected in
Table~\ref{tab:finalz}.

At several points in Tables~\ref{tab:570zf}-~\ref{tab:640zf}, combined
fits including several nearby values of $n$ and $s$ yield large $\chi^2$
while separate fits to smaller subsets of $n$ and $s$ give nearly equal
masses and acceptable $\chi^2$. This phenomenon, we have found, does not
indicate a problem with our data or our fits and arises instead because
propagators with nearby values of $n$ and $s$ in some cases are very
highly correlated and yield slightly different masses. A similar problem
would arise in trying to fit a single value $x$ to, say, a gaussian
random variable $X$ with dispersion 1, and a shifted copy $X + 0.0001$.
For any choice of $x$ the fit's $\chi^2$ is infinite.  Nonetheless, for
a Monte Carlo ensemble of 1000 $X$ values, taking $x$ as either $<X> \pm
1/\sqrt{1000}$ or $<X> + 0.0001 \pm 1/\sqrt{1000}$ is a reliable
estimate of the mean of $X$ with systematic error much smaller than the
statistical error.

An alternative way to extract a single mass from glueball
propagators for a range of $n$, $\epsilon$ and $s$ uses the
matrix of propagators
\begin{eqnarray}
\label{defMS}
M^{k \delta r n \epsilon s}_S( t_1 - t_2 ) =  
< S^{k \delta r}( t_1) S^{n \epsilon s}( t_2) > 
- < S^{k \delta r}( t_1)> <S^{n \epsilon s}( t_2) >.
\end{eqnarray}
For large $t$ and lattice time direction period $L$, $M^{k
\delta r n \epsilon s}_S( t )$ has the asymptotic form
\begin{eqnarray}
\label{Masym}
M^{k \delta r n \epsilon s}_S( t ) \rightarrow Z^{k \delta r n \epsilon
s} \{ exp( -m t) + exp[ -m (L - t)] \}
\end{eqnarray}
where $m$ is the mass of the lightest scalar glueball and $Z^{k \delta r
n \epsilon s}$ is a matrix independent of $t$.  In principle, $M^{k
\delta r n \epsilon s}_S( t )$ can be extracted from our data and fitted
to Eq.~(\ref{Masym}) to produce a value for $m$. To find the best $m$
and $Z^{k \delta r n \epsilon s}$ by minimizing the fit's $\chi^2$,
however, requires the statistical correlation matrix among the fitted
$M^{k \delta r n \epsilon s}_S( t )$.  If we fit, for example, to three
choices of $( k, \delta, r)$, three choices of $(n, \epsilon, s)$ and
four values of $t$, the correlation matrix has 1296 entries. Our
underlying data set is too small to provide reliable entries
for such a large correlation matrix. As a consequence the value of $m$
determined this way will have a statistical error which can not be
estimated reliably.  In practice, we found that the value of $m$ produced by
this method was not stabile as we varied the sets of $(m, \delta, r)$
and $(n, \epsilon, s)$ and the range of $t$ used in the fit.

\section{Tensor Propagators and Masses}\label{sect:tensor}

A propagator for tensors is defined to be
\begin{eqnarray}
\label{defPT}
P^{n \epsilon s}_T( t_1 - t_2 )  =  
\sum_i [< T_i^{n \epsilon s}( t_1) T_i^{n \epsilon s}( t_2) > 
- < T_i^{n \epsilon s}( t_1)> <T_i^{n \epsilon s}( t_2) >].
\end{eqnarray}
where $T_1$ and $T_2$ are the tensor glueball operators of
Eq.~(\ref{deftensor}), and $P^{n \epsilon s}_T( t_1 - t_2 )$ is then
averaged over reflections and time direction displacements of $t_1$ and
$t_2$ to reduce statistical noise. 

Tensor propagators were found for gauge configuration ensembles and
operator parameters listed in Tables~\ref{tab:lats} and
\ref{tab:nepsilons}.  A tensor glueball mass was extracted from 
propagators by fitting the data to the tensor version of
Eq.~(\ref{scalarasym}). We obtained a satisfactory tensor glueball mass
signal only for the lattices with $\beta$ of 5.93, 6.17 and 6.40. We did
not find an acceptable tensor signal at $\beta$ of 5.70. Overall, the
statistical errors in the tensor data are larger than those in the
scalar data of Section~\ref{sect:scalar} and, as a result, the fitting
process encounters complications not present in the scalar fits.

Tables~\ref{tab:593st} - \ref{tab:640t} list tensor masses for each
gauge ensemble with $\beta$ of 5.93 and above, for each set of operator
parameters in Table~\ref{tab:nepsilons}, fitted on one or, in some
cases, two choices of time interval.  For all fits the high end of the
fitting range $t_{max}$ is chosen to be the largest value at which a
statistically significant effective mass is found. The low end of the
fitting range $t_{min}$ is then progressively increased. The smallest
$t_{min}$ yielding a mass within one standard deviation of the masses
for all larger $t_{min}$ is selected as the lower bound for an initial
choice of the fitting range. For the lattice $16^3 \times 24$ at
$\beta$ of 5.93 and for the lattice $32^2 \times 30 \times 40$ and
$\beta$ of 6.40, however, we found that for almost all choices of
operator parameters a $t_{min}$ one unit larger than the initial choice
yielded a noticeably lower mass. These second values of $t_{min}$ and
the corresponding masses are also listed in Tables~\ref{tab:593lt}, and
\ref{tab:640t}.

\begin{figure}
\epsfxsize=\textwidth
\epsfbox{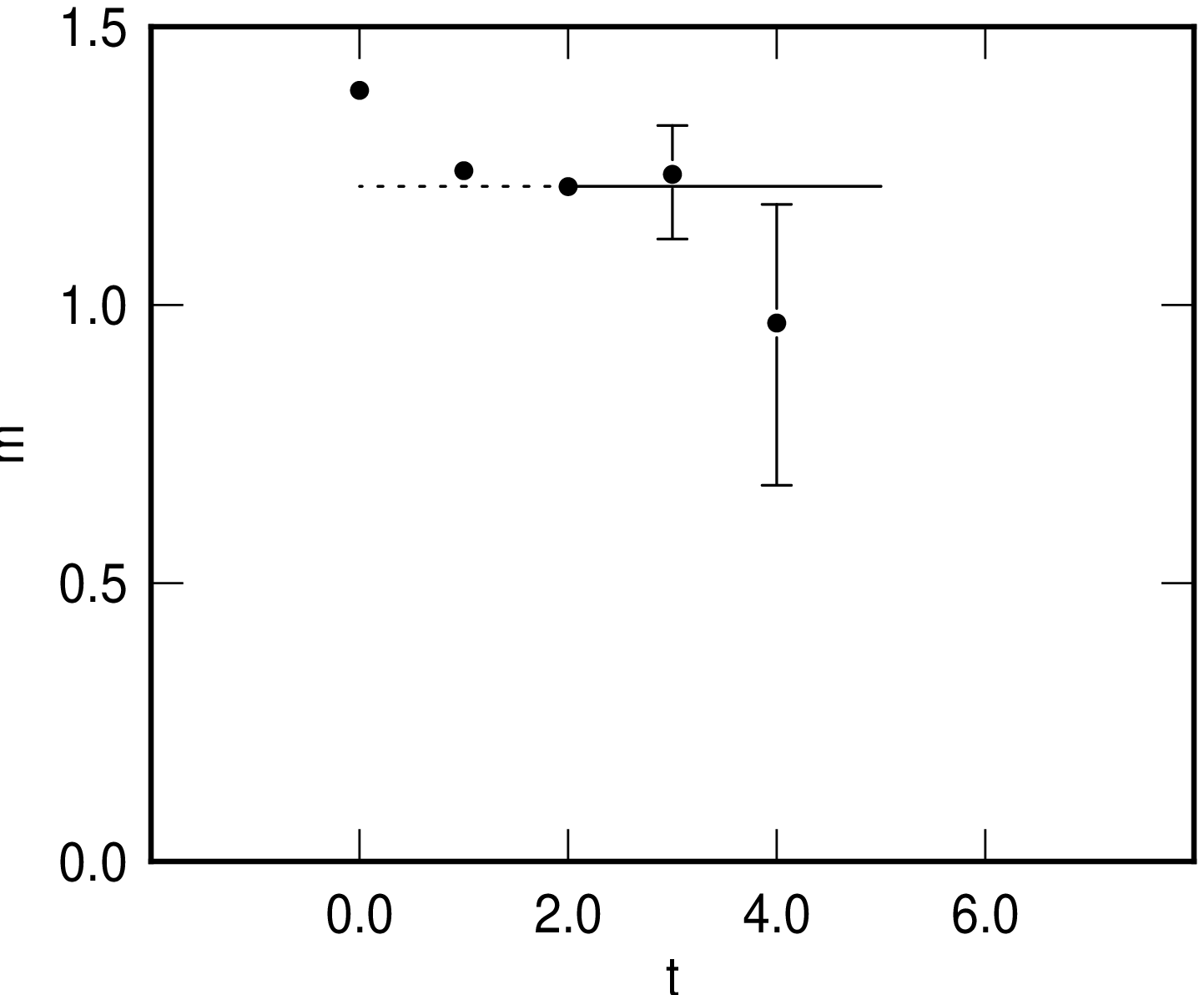}
\caption{Tensor mass fit and
tensor effective mass as a function of $t$ for the
lattice $12^3 \times 24$ at $\beta$ of 5.93 using the smeared operator
with $n$ of 6, $\epsilon$ of 1.0 and $s$ of 6.}
\label{fig:tx12b593n6s6}
\end{figure}

\begin{figure}
\epsfxsize=\textwidth
\epsfbox{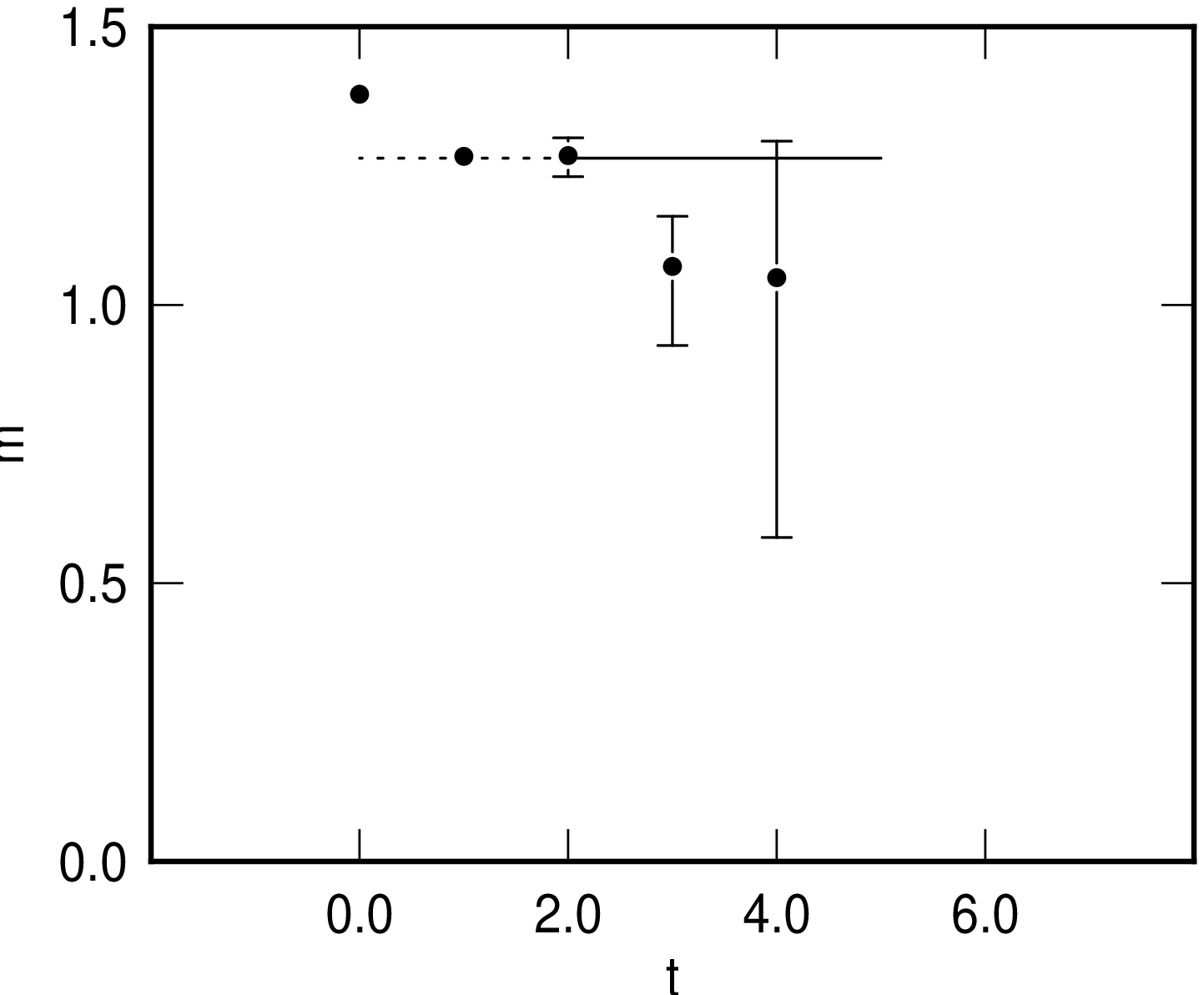}
\caption{Tensor mass fit and
tensor effective mass as a function of $t$ for the
lattice $16^3 \times 24$ at $\beta$ of 5.93 using the smeared operator
with $n$ of 6, $\epsilon$ of 1.0 and $s$ of 5.}
\label{fig:tx16b593n7s5}
\end{figure}

\begin{figure}
\epsfxsize=\textwidth
\epsfbox{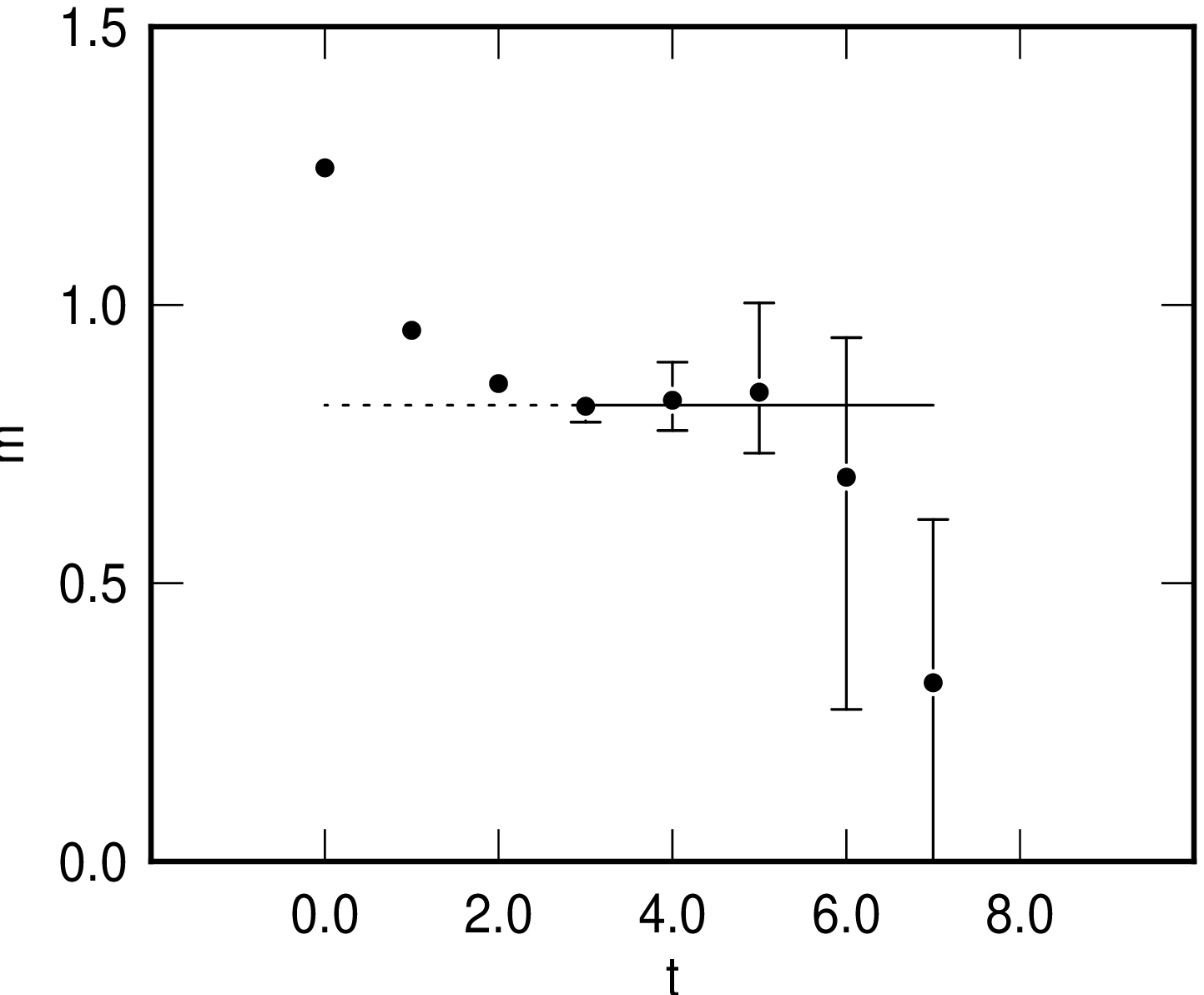}
\caption{Tensor mass fit and
tensor effective mass as a function of $t$ for the
lattice $24^3 \times 36$ at $\beta$ of 6.17 using the smeared operator
with $n$ of 7, $\epsilon$ of 1.0 and $s$ of 9.}
\label{fig:tx24b617n7s8}
\end{figure}

\begin{figure}
\epsfxsize=\textwidth
\epsfbox{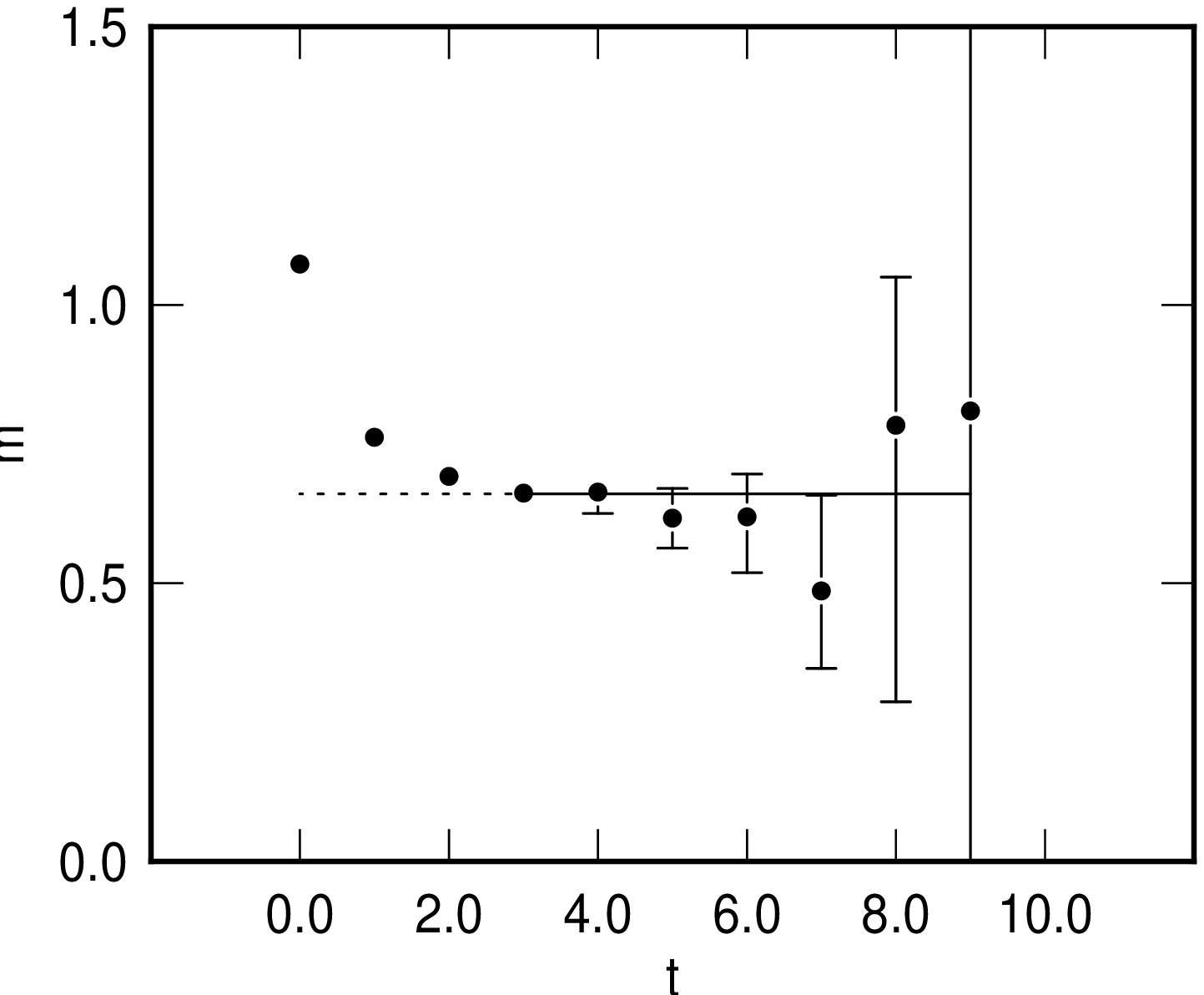}
\caption{Tensor mass fit and
tensor effective mass as a function of $t$ for the
lattice $32^2 \times 30 \times 40$ at $\beta$ of 6.40 using the smeared operator
with $n$ of 8, $\epsilon$ of 1.0 and $s$ of 10.}
\label{fig:tx32b640n8s10}
\end{figure}

Effective mass plots for tensors are shown in
Figures~\ref{fig:tx12b593n6s6} - \ref{fig:tx32b640n8s10}, for the four
lattices with $\beta$ of 5.93 and larger, for typical choices of
operator parameters. The solid line in each figure indicates the mass
obtained from a fit over the time interval which the line spans.  The
dashed line in each figures extend the solid line to smaller $t$ to show
the approach of effective masses to the fitted value.

Tables~\ref{tab:593stf} - \ref{tab:640tf} list tensor masses found by
combining, as discussed in Section~\ref{sect:scalar}, the masses fitted
to various sets of operators and choices of time interval.
Table~\ref{tab:593stf} corresponds to the lattices $12^3 \times 24$ at
$\beta$ of 5.93 with fits using the single time interval given in
Table~\ref{tab:593stf}.  Table~\ref{tab:617tf} corresponds to the lattice
$24^3 \times 36$ at $\beta$ of 6.17 with fits using the single time
interval in Table~\ref{tab:617t}.  In Tables~\ref{tab:593stf} and
\ref{tab:617tf}, all combined fits with acceptable $\chi^2$ per degree
of freedom give masses consistent with each other to within statistical
uncertainties.  In each case, the mass corresponding to the largest set
with acceptable $\chi^2$, marked with an arrow, is chosen as the final
value.

Table~\ref{tab:593ltf} for the lattice $16^3 \times 24$ at $\beta$ of
5.93 shows combined fits using both choices of $t_{min}$ of
Table~\ref{tab:593lt}.  The combined fits using the smaller $t_{min}$
have unacceptably high $\chi^2$ per degree of freedom. For the fits
using the larger $t_{min}$ the $\chi^2$ is acceptable, and the fitted
masses are all consistent with each other within statistical
uncertainties. The mass for the largest set of operators with the larger
$t_{min}$ is chosen as the final number.  Table~\ref{tab:640tf} for the
lattice $32^2 \times 30 \times 40$ at $\beta$ of 6.40 also gives
combined fits for both $t_{min}$ in Table~\ref{tab:640t}.  Most fits for
both $t_{min}$ have acceptable $\chi^2$ per degree of freedom.  The
masses obtained from the larger $t_{min}$ all lie one standard deviation
or a bit more below the masses found with the smaller $t_{min}$,
however, and all have significantly better $\chi^2$ than the fits with
the smaller $t_{min}$.  The mass found from the largest set of operators
for the larger $t_{min}$ is therefore chosen as the final result.

The collection of final tensor masses is listed in
Table~\ref{tab:finalt}.

\section{Volume Dependence}\label{sect:vol}

We now consider an estimate of the difference between the scalar and
tensor glueball masses in Table~\ref{tab:finalz} and \ref{tab:finalt}
for finite lattice period $L$ and the infinite volume limits of these
quantities.

For large values of $L$, scalar $m_0(L)$ and tensor
$m_2(L)$ glueball masses deviate from their infinite volume limits,
$m_0$ and $m_2$, respectively, by~\cite{Luscher}
\begin{eqnarray}
\label{luscher}
m_s( L) = m_s
\{ 1 - g_s \frac{ exp(\frac{ -\sqrt{3} m_0 L}{2}) } {m_0 L} 
- O[ \frac{ exp( -m_0 L)}{m_0 L}] \}
\end{eqnarray}
where $s$ is 0 or 2. In Ref.~\cite{Schierholz} for $\beta$ near 6.0, data
for $m_0(L)$ is shown to fit the two leading terms in
Eq.~(\ref{luscher}) reasonably well at 4 values of $L$ ranging from
$6/m_0$ to $12/m_0$.  This result is plausible since for $L$ ranging
from $6/m_0$ to $12/m_0$, the third term in Eq.~(\ref{luscher}) is
smaller than the second by a factor ranging from $O(0.4)$ to
$O(0.2)$. For our data at $\beta$ of 5.93, Table~\ref{tab:finalz}
shows that $m_0$ is above 0.75 so that $L$ of 12 and 16 are larger than
$8/m_0$ and $12/m_0$, respectively.  Thus we believe that for the data
at $\beta$ of 5.93, the leading two terms of Eq.~(\ref{luscher}) likely
provide a fairly reliable estimate of the $L$ dependence of $m_0(L)$ and
$m_2(L)$.

Fitting the $\beta=5.93$ data in Table~\ref{tab:finalz} to the two
leading terms of Eq.~(\ref{luscher}) yields $m_0$ of $0.783 \pm 0.012$
and $g_0$ of $1500 \pm 1100$.  In addition a bootstrap calculation
yields with 95\% probability
\begin{eqnarray}
\label{volerr0}
\frac{ m_0 - m_0(16)}{m_0} \le 0.0037.
\end{eqnarray}.
At $\beta=5.93$, Table~\ref{tab:finalt} combined with the leading two
terms of Eq.~(\ref{luscher}) gives $m_2$ of $1.236 \pm 0.037$ and $g_2$
is $1300 \pm 1200$. A bootstrap calculation yields with 95\% probability
\begin{eqnarray}
\label{volerr2}
\frac{ m_2 - m_2(16)}{m_2} \le 0.0048
\end{eqnarray}.

Overall, it appears to us safe to conclude that at $\beta$ of 5.93 the
difference between scalar and tensor masses for $L$ of 16 and their
infinite volume limits are of the order of 0.5\% or less.  In
Section~\ref{sect:cont} we show that the scalar and tensor glueball
masses in Tables~\ref{tab:finalz} and \ref{tab:finalt} with $\beta$ of
5.93 and greater and $m_0 L$ fixed at about 13 are not far from
asymptotic scaling. We therefore expect the fractional volume dependent
errors found in these masses to be about the same as the errors at
$\beta$ of 5.93. Thus the finite volume errors in all masses in
Tables~\ref{tab:finalz} and \ref{tab:finalt} with $\beta$ of 5.93 and
greater and $m_0 L$ of about 13 should be 0.5\% or less.

\section{Continuum Limit}\label{sect:cont}

The nonzero lattice spacing scalar and tensor glueball masses in lattice
units given in Tables~\ref{tab:finalz} and \ref{tab:finalt},
respectively, we now convert to physical units and extrapolate to zero
lattice spacing.

To convert masses in lattice units to physical units, we divide by a
known mass measured in lattice units.  One natural choice for this
conversion factor is the rho mass $m_{\rho}(a) a$. Values of
$m_{\rho}(a) a$ for three of the four $\beta$ in Tables~\ref{tab:finalz}
and \ref{tab:finalt} are given in Ref.~\cite{hadrons}. For the largest
$\beta$ in Tables~\ref{tab:finalz} and \ref{tab:finalt},
Ref.~\cite{hadrons} does not report $m_{\rho}(a) a$.  For the three
$\beta$ considered in Ref.~\cite{hadrons}, however, the ratio
$[\Lambda^{(0)}_{\overline{MS}} a]/[m_{\rho}(a) a]$ is found to be
independent of $\beta$ to within statistical errors. Here
$\Lambda^{(0)}_{\overline{MS}} a$ is obtained by the 2-loop
Callan-Symanzik equation from $\alpha_{\overline{MS}}$ found from its
mean-field improved \cite{Lepage} relation to $\beta$. Since
$[\Lambda^{(0)}_{\overline{MS}} a]/[m_{\rho}(a) a]$ is constant within
errors, converting to physical units using
$\Lambda^{(0)}_{\overline{MS}} a$ then extrapolating to zero lattice
spacing should give results nearly equivalent to those found using
$m_{\rho}(a) a$.  Table~\ref{tab:lambdas} lists, for each $\beta$, the
corresponding mean-field improved $\alpha_{\overline{MS}}$ and
$\Lambda^{(0)}_{\overline{MS}} a$.

The $\beta$ dependence of valence approximation glueball masses is
determined entirely by the pure gauge part of the QCD action.  The
leading irrelevant operator in the pure gauge plaquette action has
lattice spacing dependence of $O(a^2)$. Thus for scalar and tensor
glueball masses $m_0$ and $m_2$, respectively, we extrapolate to the
continuum limit by
\begin{eqnarray}
\label{contlim}
\frac{ m_s(a) a}{ \Lambda^{(0)}_{\overline{MS}} a} & = &
 \frac{ m_s}{ \Lambda^{(0)}_{\overline{MS}}} + 
 C [\Lambda^{(0)}_{\overline{MS}} a]^2,
\end{eqnarray}
where $s$ is 0 or 2.  

If $\Lambda^{(0)}_{\overline{MS}} a$ in Eq.~(\ref{contlim}) were
replaced by $m_{\rho}(a) a$, then since the leading irrelevant operator
in the quark action has lattice spacing dependence of $O(a)$ it might be
argued that the quadratic $O(a^2)$ term in the equation's right hand
side should be a linear $O(a)$. This in turn would contradict our claim
that extrapolation using either $m_{\rho}(a) a$ or
$\Lambda^{(0)}_{\overline{MS}} a$ will give nearly equal results. An
answer to this objection is that the approximate constancy of
$[\Lambda^{(0)}_{\overline{MS}} a]/[m_{\rho}(a) a]$ implies that the
$O(a)$ irrelevant contribution to $m_{\rho}(a) a$ is quite small.  The
constancy of $[\Lambda^{(0)}_{\overline{MS}} a]/[m_{\rho}(a) a]$ as a
function of $a$ or equivalently as a function of $\beta$ can not be
explained by a cancellation of an $O(a)$ term in
$\Lambda^{(0)}_{\overline{MS}} a$ with an $O(a)$ term in $m_{\rho}(a) a$
since $\Lambda^{(0)}_{\overline{MS}} a$ is defined to fulfill the true
continuum two-loop Callan-Synamzik equation and itself has no $O(a)$
corrections.  The leading correction to the $\beta$ dependence of
$\Lambda^{(0)}_{\overline{MS}} a$ is by a multiplicative factor of $[1 +
O(\beta^2)]$.  If $\Lambda^{(0)}_{\overline{MS}} a$ is replaced by
$m_{\rho}(a) a$, any significant $a$ dependence which appears will come
from the $O(a^2)$ term in $m_s(a) a$.  Thus Eq.~(\ref{contlim}) even
with $m_{\rho}(a) a$ substituted for $\Lambda^{(0)}_{\overline{MS}} a$
will remain correct.

\begin{figure}
\epsfxsize=\textwidth
\epsfbox{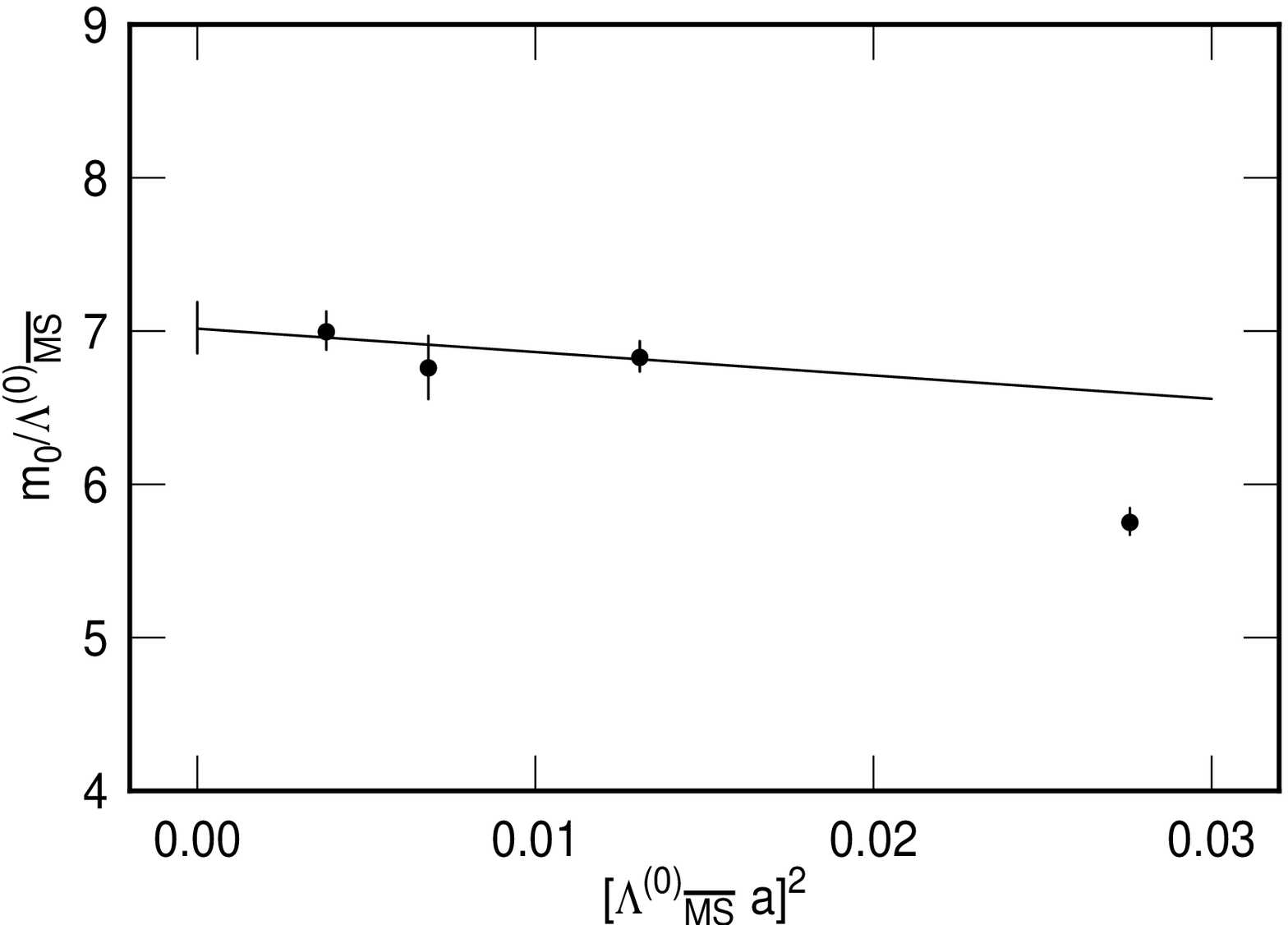}
\caption{The scalar glueball mass in units of $\Lambda^{(0)}_{\overline{MS}}$
extrapolated to zero lattice spacing linearly in $[\Lambda^{(0)}_{\overline{MS}}]^2$.}
\label{fig:scalarquad}
\end{figure}

The scalar data of Tables~\ref{tab:finalz} combined with
$\Lambda^{(0)}_{\overline{MS}} a$ of Table~\ref{tab:lambdas} fitted to
Eq.~(\ref{contlim}) at the three largest $\beta$ is shown in
Figure~\ref{fig:scalarquad}. The predicted continuum limit $m_0 /
\Lambda^{(0)}_{\overline{MS}}$ is $7.016 \pm 0.167$.  The fit in
Figure~\ref{fig:scalarquad} has a $\chi^2$ of 0.6 over a range in which 
the term $[\Lambda^{(0)}_{\overline{MS}} a]^2$ varies by more
than a factor of 3.4. The variation of $[m_s(a)
a]/[ \Lambda^{(0)}_{\overline{MS}} a]$ over the fitting range, however, is only
slight.  Each of the three nonzero lattice spacing values of $[m_s(a)
a]/[ \Lambda^{(0)}_{\overline{MS}} a]$ is within 1.6 standard deviations
of the extrapolated zero lattice spacing result.  Thus we believe the
extrapolation to zero lattice spacing is quite reliable and would expect
the predicted continuum mass to be not very different from what would be
obtained by any other reasonable, smooth extrapolation of the data.  

\begin{figure}
\epsfxsize=\textwidth
\epsfbox{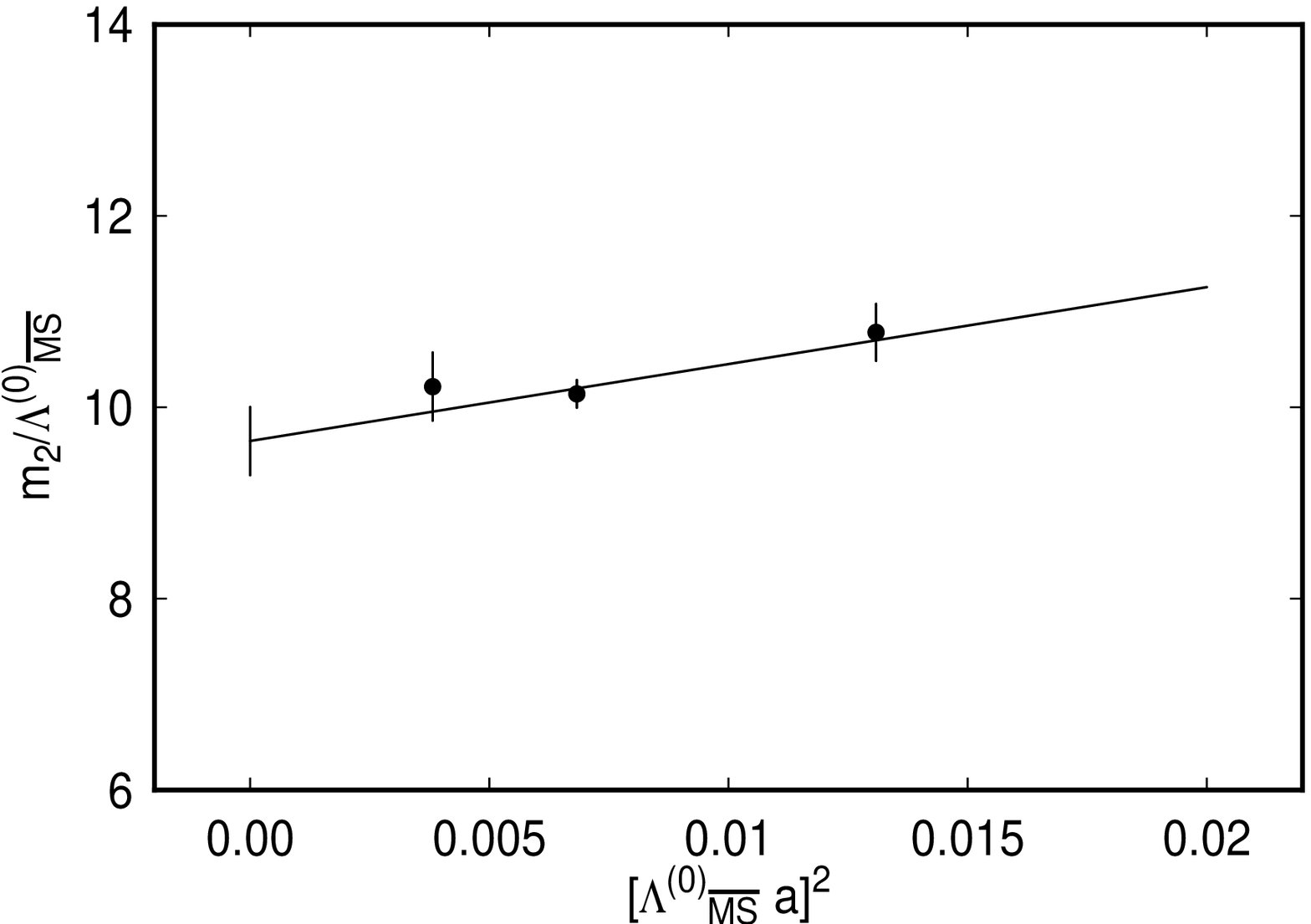}
\caption{The tensor glueball mass in units of $\Lambda^{(0)}_{\overline{MS}}$
extrapolated to zero lattice spacing linearly in $[\Lambda^{(0)}_{\overline{MS}}]^2$.}
\label{fig:tensorquad}
\end{figure}

The tensor data of Tables~\ref{tab:finalt} combined with
$\Lambda^{(0)}_{\overline{MS}} a$ of Table~\ref{tab:lambdas} fitted to
Eq.~(\ref{contlim}) at the three largest $\beta$, the only $\beta$ for which
tensor masses were found, is shown in
Figure~\ref{fig:tensorquad}. The predicted continuum limit $m_2
/\Lambda^{(0)}_{\overline{MS}}$ is $9.65 \pm 0.36$.  The fit in
Figure~\ref{fig:tensorquad} has a $\chi^2$ of 0.8, while,
as before, 
the term $[\Lambda^{(0)}_{\overline{MS}} a]^2$ in Eq.~(\ref{contlim})
varies by more than a factor of 3.4 over the fitting range. 

To obtain scalar and tensor glueball masses in units of MeV, we combine
the continuum limit $\Lambda^{(0)}_{\overline{MS}}/m_{\rho}$ of $0.305
\pm 0.008$ \cite{hadrons} with $m_{\rho}$ of 770 MeV to give
$\Lambda^{(0)}_{\overline{MS}}$ of $234.9 \pm 6.2$ MeV.  The scalar
glueball mass becomes $1648 \pm 58$ MeV and the tensor mass becomes
$2267 \pm 104$ MeV. The continuum limit results are summarized in
Table~\ref{tab:contlim}.

For $\Lambda^{(0)}_{\overline{MS}}/m_{\rho}$ we take the value given in
Ref.~\cite{hadrons} for a lattice with period of about 2.4 fermi. For
the rho mass obtained at $\beta$ of 5.7 from a combination of
propagators for rho operators with smearing parameters 0, 1 and 2, the
2.4 fermi result differs from the result for period 3.6 fermi by a bit
over one standard deviation.  This difference appears to be largely a
consequence of a slightly poorer separtion of the rho component of the
propagator from excited state components in the 2.4 fermi rho mass
calculation than in the 3.6 fermi calculation~\cite{Gottlieb}.  For the
rho operator with smearing parameter 4, which couples more weakly to
excited states, the difference at $\beta$ of 5.7 between 2.4 fermi and
3.6 fermi predictions is much less than one standard deviation. Thus
overall it appears to us reasonable to take the 2.4 fermi calculations
as the infinite volume limit, within statistical errors. The continuum
limit values of $\Lambda^{(0)}_{\overline{MS}}/m_{\rho}$ for the the
data combining smearings 0, 1 and 2 and for the data from smearing 4 are
nearly identical.

\section{Comparison with Other Results}\label{sect:comp}

An independent calculation of the infinite volume, continuum limit of
the valence approximation to several glueball masses is reported in
Ref.~\cite{livertal}. A second, more recent, calculation appears in
Ref.~\cite{Morningstar}. A comparison of Ref.~\cite{livertal} with the original
analysis~\cite{masses} of our results appears in Ref.~\cite{review}.

The calculation of Ref.~\cite{livertal} uses the same plaquette action
we use but takes a different set of glueball operators.  The gauge field
ensembles of Ref.~\cite{livertal} range from 1000 to 3000
configurations.  For the scalar and tensor masses Ref.~\cite{livertal}
reports $1550 \pm 50$ MeV and $2270 \pm 100$ MeV, respectively.  The
predicted zero lattice spacing masses are not actually found by
extrapolation to zero lattice spacing, but are obtained instead from
calculations at $\beta$ of 6.40 of glueball masses in units of the
square root of string tension, $\sqrt{\sigma}$, then converted to MeV
using an assumed $\sqrt{\sigma}$ of 440 MeV with zero uncertainty. The
uncertainties given in the masses are entirely the uncertainties in the
$\beta$ of 6.40 calculations of masses in units of $\sqrt{\sigma}$ and
are thus missing at least a contribution from the uncertainty in
$\sqrt{\sigma}$. A graph shown in Ref.~\cite{livertal} suggests that the
$\beta$ of 6.40 value of $[m_0( a) a]/[\sqrt{\sigma (a)} a]$ is about 50
MeV below the data's zero lattice spacing limit.  An additional error of
$\pm 50$ MeV in the scalar mass is therefore proposed in
Ref.~\cite{livertal} as a consequence of the absence of extrapolation to
zero lattice spacing. Since $[m_0( a) a]/[\sqrt{\sigma (a)} a]$ of
Ref.~\cite{livertal} is clearly rising as lattice spacing falls, it does
not appear to us that a symmetric error of $\pm 50$ MeV an accurate 
representation of the effect of the absence of extrapolation.  If the
statistical error and extrapolation error in the scalar mass are,
nonetheless, taken at face value and combined the result is a prediction
of $1550 \pm 71$ MeV.  No estimate is given for the extrapolation error
in the tensor mass, which is found to be only weakly dependent on
lattice spacing if measured in units of $\sqrt{\sigma}$.  A scalar mass
of $1550 \pm 71$ MeV is a bit over one standard deviations below the
result $1648 \pm 58$ MeV in Table~\ref{tab:contlim}, while the tensor
mass of $2270 \pm 100$ MeV is in close agreement with our value of $2267
\pm 104$ MeV.

If the continuum limit of the Ref.~\cite{livertal} data is found by
extrapolation to zero lattice spacing of $[m_0( a)
a]/[\Lambda^{(0)}_{\overline{MS}} a]$, following Section~\ref{sect:cont},
the result for $m_0/\Lambda^{(0)}_{\overline{MS}}$ is $6.67 \pm 0.33$.
Converted to MeV using $\Lambda^{(0)}_{\overline{MS}}$ of $234.9 \pm
6.2$ MeV, $m_0$ becomes $1567 \pm 88$ MeV. This value is less than a
standard deviation below the prediction $1648 \pm 58$ MeV in
Table~\ref{tab:contlim}.

The calculation of Ref.~\cite{Morningstar} uses an improved action with
time direction lattice spacing chosen smaller than the space direction.
The gauge field ensembles range in size from 4000 to 10000
configurations.  Masses measured in units of the parameter
$r_0^{-1}$~\cite{Sommer} are extrapolated to zero lattice spacing, then
converted to MeV using a value of $r_0^{-1}$ found by extrapolation of
$r_0^{-1}/m_{\rho}$ to zero lattice spacing.  As a result of working at
relatively large values of lattice spacing, some ambiguity is
encountered in matching the scalar mass's lattice spacing dependence to
the small lattice spacing asymptotic behavior expected for the improved
action. Taking this uncertainty into account, the scalar mass is
predicted to be $1730 \pm 94$ MeV. The tensor mass, for which the
extrapolation to zero lattice spacing encounters no problem, is
predicted to be $2400 \pm 122$ MeV.  Both numbers are a bit under one
standard deviation above the predictions in Table~\ref{tab:contlim}.
For the ratio $m_2/m_0$ Ref.~\cite{Morningstar} predicts $1.39
\pm 0.04$, in good agreement with the value $1.375 \pm
0.066$ in Table~\ref{tab:contlim}. Thus the difference between
Table~\ref{tab:contlim} and Ref.~\cite{Morningstar} is almost entirely a
discrepancy in overall mass scale.

Combining our extrapolation of $6.67 \pm 0.33$ for the data in
Ref.~\cite{livertal} with $7.016 \pm 0.167$ in Table~\ref{tab:contlim}
gives $6.95 \pm 0.15$ for $m_0/\Lambda^{(0)}_{\overline{MS}}$, thus
$1631 \pm 55$ MeV.  Combining $1631 \pm 55$ MeV with $1730 \pm 94$ MeV
of Ref.~\cite{Morningstar} gives a world average valence approximation
scalar mass of $1656 \pm 47$ MeV. This number is consistent with the
unmixed scalar mass of $1622 \pm 29$ MeV found in Ref.~\cite{lat98mix}
taking the observed states $f_0(1710)$, $f_0(1500)$ and $f_0(1400)$ as
the mixed versions of the scalar glueball and the two isoscalar spin
zero quarkonium states, respectively. The state $f_0(1710)$ in this
calculation is assigned a glueball component of $73.8 \pm 9.5$ \%.
Combining $2270 \pm 100$ MeV, $2267 \pm 104$ MeV and $2400 \pm 122$ MeV
gives a world average valence approximation tensor mass of $2302 \pm 62$
MeV.

\begin{table}
\begin{center}
\begin{tabular}{ccccc}     \hline
 $\beta$ &  lattice         & skip & count & bin \\ \hline
 5.70    & $16^3 \times 24$ & 50 & 8,094  & 8   \\ 
 5.93    & $12^3 \times 24$ & 25 & 48,278 & 16  \\
         & $16^3 \times 24$ & 25 & 30,640 & 16  \\
 6.17    & $24^3 \times 36$ & 25 & 31,150 & 16  \\
 6.40    & $30 \times 32^2 \times 40$ 
                            & 25 & 25,440 & 16  \\  \hline
\end{tabular}
\caption{Configurations analyzed.}
\label{tab:lats}
\end{center}
\end{table}

\begin{table}
\begin{center}
\begin{tabular}{ccccc}     \hline
 $\beta$ &  lattice         &  n & $\epsilon$ & s \\ \hline
 5.70    & $16^3 \times 24$ &  $3-10$& 0.25 & $1-4$  \\ 
 5.93    & $12^3 \times 24$ &  $5-7$ & 1.00 & $4-6$  \\
         & $16^3 \times 24$ &  $5-8$ & 1.00 & $3-7$  \\
 6.17    & $24^3 \times 36$ &  7,8   & 1.00 & $7-10$ \\
 6.40    & $30 \times 32^2 \times 40$ 
                            &  6,8   & 1.00 & $7-11$ \\  \hline
\end{tabular}
\caption{Glueball operator parameters.}
\label{tab:nepsilons}
\end{center}
\end{table}

\begin{table}
\begin{center}
\begin{tabular}{cccccc}     \hline
$n$ & $s$ & $t_{min}$ & $t_{max}$ & mass & $\chi^2/d.o.f.$ \\ \hline

 5 & 4 & 1 & 8 & $0.827 \pm 0.006$  & 2.26 \\ \hline 
 5 & 4 & 2 & 8 & $0.792 \pm 0.013$  & 1.14 \\ \hline 
 5 & 4 & 3 & 8 & $0.817 \pm 0.029$  & 1.22 \\ \hline 
 5 & 4 & 4 & 8 & $0.831 \pm 0.068$  & 1.61 \\ \hline 
\hline
 5 & 5 & 1 & 8 & $0.810 \pm 0.006$  & 0.85  \\ \hline 
 5 & 5 & 2 & 8 & $0.791 \pm 0.013$  & 0.55  \\ \hline 
 5 & 5 & 3 & 8 & $0.798 \pm 0.028$  & 0.67  \\ \hline 
 5 & 5 & 4 & 8 & $0.767 \pm 0.059$  & 0.80  \\ \hline 
\hline
 5 & 6 & 1 & 8 & $0.824 \pm 0.007$  & 1.32  \\ \hline 
 5 & 6 & 2 & 8 & $0.793 \pm 0.014$  & 0.58  \\ \hline 
 5 & 6 & 3 & 8 & $0.785 \pm 0.031$  & 0.71  \\ \hline 
 5 & 6 & 4 & 8 & $0.721 \pm 0.066$  & 0.57  \\ \hline 
\end{tabular}
\caption{Fitted scalar glueball mass in lattice units for various choices
of $n$, $s$ and fitting range for the lattice $16^3 \times 24$ at $\beta = 5.93$.}
\label{tab:593zd1}
\end{center}
\end{table}
 
\begin{table}
\begin{center}
\begin{tabular}{cccccc}     \hline
$n$ & $s$ & $t_{min}$ & $t_{max}$ & mass & $\chi^2/d.o.f.$ \\ \hline

 6 & 4 & 1 & 8 & $0.815 \pm 0.006$  & 1.82  \\ \hline 
 6 & 4 & 2 & 8 & $0.785 \pm 0.012$  & 0.96  \\ \hline 
 6 & 4 & 3 & 8 & $0.813 \pm 0.027$  & 0.90  \\ \hline 
 6 & 4 & 4 & 8 & $0.806 \pm 0.064$  & 1.20  \\ \hline 
\hline
 6 & 5 & 1 & 8 & $0.800 \pm 0.006$  & 0.70  \\ \hline 
 6 & 5 & 2 & 8 & $0.785 \pm 0.012$  & 0.50  \\ \hline 
 6 & 5 & 3 & 8 & $0.799 \pm 0.027$  & 0.55  \\ \hline 
 6 & 5 & 4 & 8 & $0.755 \pm 0.054$  & 0.53  \\ \hline 
\hline
 6 & 6 & 1 & 8 & $0.811 \pm 0.006$  & 0.96  \\ \hline 
 6 & 6 & 2 & 8 & $0.788 \pm 0.013$  & 0.51  \\ \hline 
 6 & 6 & 3 & 8 & $0.789 \pm 0.028$  & 0.64  \\ \hline 
 6 & 6 & 4 & 8 & $0.722 \pm 0.060$  & 0.39  \\ \hline 
\end{tabular}
\caption{Fitted scalar glueball mass in lattice units for various choices
of $n$, $s$ and fitting range for the lattice $16^3 \times 24$ at $\beta = 5.93$.}
\label{tab:593zd2}
\end{center}
\end{table}

\begin{table}
\begin{center}
\begin{tabular}{cccccc}     \hline
$n$ & $s$ & $t_{min}$ & $t_{max}$ & mass & $\chi^2/d.o.f.$ \\ \hline

 7 & 4 & 1 & 8 & $0.803 \pm 0.006$  & 1.41  \\ \hline 
 7 & 4 & 2 & 8 & $0.777 \pm 0.012$  & 0.70  \\ \hline 
 7 & 4 & 3 & 8 & $0.809 \pm 0.026$  & 0.48  \\ \hline 
 7 & 4 & 4 & 8 & $0.782 \pm 0.059$  & 0.58  \\ \hline 
\hline
 7 & 5 & 1 & 8 & $0.792 \pm 0.006$  & 0.61  \\ \hline 
 7 & 5 & 2 & 8 & $0.779 \pm 0.012$  & 0.45  \\ \hline 
 7 & 5 & 3 & 8 & $0.798 \pm 0.027$  & 0.42  \\ \hline 
 7 & 5 & 4 & 8 & $0.746 \pm 0.052$  & 0.25  \\ \hline 
\hline
 7 & 6 & 1 & 8 & $0.800 \pm 0.006$  & 0.75  \\ \hline 
 7 & 6 & 2 & 8 & $0.783 \pm 0.013$  & 0.47  \\ \hline 
 7 & 6 & 3 & 8 & $0.793 \pm 0.027$  & 0.55  \\ \hline 
 7 & 6 & 4 & 8 & $0.723 \pm 0.056$  & 0.20  \\ \hline 
\end{tabular}
\caption{Fitted scalar glueball mass in lattice units for various choices
of $n$, $s$ and fitting range for the lattice $16^3 \times 24$ at $\beta = 5.93$.}
\label{tab:593zd3}
\end{center}
\end{table}
 
\begin{table}
\begin{center}
\begin{tabular}{cccccc}     \hline
$n$ & $s$ & $t_{min}$ & $t_{max}$ & mass & $\chi^2/d.o.f.$ \\ \hline

 8 & 4 & 1 & 8 & $0.789 \pm 0.007$  & 0.97  \\ \hline 
 8 & 4 & 2 & 8 & $0.765 \pm 0.013$  & 0.49  \\ \hline 
 8 & 4 & 3 & 8 & $0.803 \pm 0.031$  & 0.18  \\ \hline 
 8 & 4 & 4 & 8 & $0.754 \pm 0.063$  & 0.04  \\ \hline 
\hline
 8 & 5 & 1 & 8 & $0.784 \pm 0.006$  & 0.63  \\ \hline 
 8 & 5 & 2 & 8 & $0.768 \pm 0.012$  & 0.40  \\ \hline 
 8 & 5 & 3 & 8 & $0.791 \pm 0.029$  & 0.31  \\ \hline 
 8 & 5 & 4 & 8 & $0.748 \pm 0.055$  & 0.23  \\ \hline 
\hline
 8 & 6 & 1 & 8 & $0.791 \pm 0.006$  & 0.77  \\ \hline 
 8 & 6 & 2 & 8 & $0.775 \pm 0.013$  & 0.57  \\ \hline 
 8 & 6 & 3 & 8 & $0.797 \pm 0.028$  & 0.54  \\ \hline 
 8 & 6 & 4 & 8 & $0.716 \pm 0.057$  & 0.04  \\ \hline 
\end{tabular}
\caption{Fitted scalar glueball mass in lattice units for various choices
of $n$, $s$ and fitting range for the lattice $16^3 \times 24$ at $\beta = 5.93$.}
\label{tab:593zd4}
\end{center}
\end{table}

\begin{table}
\begin{center}
\begin{tabular}{cccccc}     \hline
$n$ & $s$ & $t_{min}$ & $t_{max}$ & mass & $\chi^2/d.o.f.$ \\ \hline

 4 & 1 & 1 & 5 & $0.971 \pm 0.019$  & 0.23  \\ \hline 
 4 & 2 & 1 & 5 & $0.959 \pm 0.017$  & 0.35  \\ \hline 
 4 & 3 & 1 & 5 & $0.952 \pm 0.016$  & 0.39  \\ \hline 
 4 & 4 & 1 & 5 & $0.984 \pm 0.023$  & 0.26  \\ \hline 
 5 & 1 & 1 & 5 & $0.964 \pm 0.018$  & 0.26  \\ \hline 
 5 & 2 & 1 & 5 & $0.956 \pm 0.017$  & 0.29  \\ \hline 
 5 & 3 & 1 & 5 & $0.953 \pm 0.017$  & 0.23  \\ \hline 
 5 & 4 & 1 & 5 & $0.983 \pm 0.020$  & 0.15  \\ \hline 
 6 & 1 & 1 & 5 & $0.958 \pm 0.017$  & 0.27  \\ \hline 
 6 & 2 & 1 & 5 & $0.954 \pm 0.017$  & 0.23  \\ \hline 
 6 & 3 & 1 & 5 & $0.956 \pm 0.017$  & 0.13  \\ \hline 
 6 & 4 & 1 & 5 & $0.985 \pm 0.020$  & 0.09  \\ \hline 
\end{tabular}

\caption{Fitted scalar glueball mass in lattice units using the best
$t_{min}$ and $t_{max}$ for various choices
of $n$ and $s$ for the lattice 
$16^3 \times 24$ at $\beta$ of $5.70$}
\label{tab:570z}
\end{center}
\end{table}

\begin{table}
\begin{center}
\begin{tabular}{cccccc}     \hline
$n$ & $s$ & $t_{min}$ & $t_{max}$ & mass & $\chi^2/d.o.f.$ \\ \hline

 5 & 4 & 3 & 7 & $0.752 \pm 0.021$  & 1.30  \\ \hline 
 5 & 5 & 3 & 7 & $0.737 \pm 0.020$  & 1.01  \\ \hline 
 5 & 6 & 3 & 7 & $0.747 \pm 0.022$  & 0.46  \\ \hline 
 6 & 4 & 3 & 7 & $0.754 \pm 0.020$  & 0.97  \\ \hline 
 6 & 5 & 3 & 7 & $0.742 \pm 0.020$  & 0.68  \\ \hline 
 6 & 6 & 2 & 7 & $0.772 \pm 0.010$  & 0.51  \\ \hline 
 6 & 6 & 3 & 7 & $0.751 \pm 0.022$  & 0.29  \\ \hline 
 7 & 4 & 3 & 7 & $0.757 \pm 0.020$  & 0.72  \\ \hline 
 7 & 5 & 3 & 7 & $0.747 \pm 0.020$  & 0.42  \\ \hline 
 7 & 6 & 2 & 7 & $0.772 \pm 0.010$  & 0.28  \\ \hline 
 7 & 6 & 3 & 7 & $0.756 \pm 0.021$  & 0.16  \\ \hline 
\end{tabular}

\caption{Fitted scalar glueball mass in lattice units using the best
$t_{min}$ and $t_{max}$ for various choices
of $n$ and $s$ for the lattice 
$12^3 \times 24$ at $\beta$ of $5.93$}
\label{tab:593sz}
\end{center}
\end{table}

\begin{table}
\begin{center}
\begin{tabular}{cccccc}     \hline
$n$ & $s$ & $t_{min}$ & $t_{max}$ & mass & $\chi^2/d.o.f.$ \\ \hline

 5 & 4 & 2 & 8 & $0.792 \pm 0.013$  & 1.14  \\ \hline 
 5 & 5 & 2 & 8 & $0.791 \pm 0.013$  & 0.55  \\ \hline 
 5 & 6 & 2 & 8 & $0.793 \pm 0.014$  & 0.58  \\ \hline 
 6 & 4 & 2 & 8 & $0.785 \pm 0.012$  & 0.96  \\ \hline 
 6 & 5 & 2 & 8 & $0.785 \pm 0.012$  & 0.50  \\ \hline 
 6 & 6 & 2 & 8 & $0.788 \pm 0.013$  & 0.51  \\ \hline 
 7 & 4 & 2 & 8 & $0.777 \pm 0.012$  & 0.70  \\ \hline 
 7 & 5 & 2 & 8 & $0.779 \pm 0.012$  & 0.45  \\ \hline 
 7 & 6 & 2 & 8 & $0.783 \pm 0.013$  & 0.47  \\ \hline 
 8 & 4 & 2 & 8 & $0.765 \pm 0.013$  & 0.49  \\ \hline 
 8 & 5 & 2 & 8 & $0.768 \pm 0.012$  & 0.40  \\ \hline 
 8 & 6 & 2 & 8 & $0.775 \pm 0.013$  & 0.57  \\ \hline 
\end{tabular}
\caption{Fitted scalar glueball mass in lattice units using the best
$t_{min}$ and $t_{max}$ for various choices
of $n$ and $s$ for the lattice 
$16^3 \times 24$ at $\beta$ of $5.93$}
\label{tab:593lz}
\end{center}
\end{table}

\begin{table}
\begin{center}
\begin{tabular}{cccccc}     \hline
$n$ & $s$ & $t_{min}$ & $t_{max}$ & mass & $\chi^2/d.o.f.$ \\ \hline

 7 & 7 & 4 & 9 & $0.570 \pm 0.018$  & 0.22  \\ \hline 
 7 & 8 & 4 & 9 & $0.561 \pm 0.020$  & 0.14  \\ \hline 
 7 & 9 & 4 & 9 & $0.554 \pm 0.024$  & 0.30  \\ \hline 
7 & 10 & 4 & 9 & $0.540 \pm 0.030$  & 0.38  \\ \hline 
 8 & 7 & 4 & 9 & $0.562 \pm 0.019$  & 0.18  \\ \hline 
 8 & 8 & 4 & 9 & $0.551 \pm 0.018$  & 0.17  \\ \hline 
 8 & 9 & 4 & 9 & $0.545 \pm 0.021$  & 0.10  \\ \hline 
8 & 10 & 4 & 9 & $0.534 \pm 0.027$  & 0.17  \\ \hline 
\end{tabular}
\caption{Fitted scalar glueball mass in lattice units using the best
$t_{min}$ and $t_{max}$ for various choices
of $n$ and $s$ for the lattice 
$24^3 \times 36$ at $\beta$ of $6.17$}
\label{tab:617z}
\end{center}
\end{table}

\begin{table}
\begin{center}
\begin{tabular}{cccccc}     \hline
$n$ & $s$ & $t_{min}$ & $t_{max}$ & mass & $\chi^2/d.o.f.$ \\ \hline

 6 & 7 & 4 & 12 & $0.461 \pm 0.013$   & 0.50  \\ \hline 
 6 & 8 & 4 & 12 & $0.446 \pm 0.012$   & 0.57  \\ \hline 
 6 & 9 & 3 & 12 & $0.448 \pm 0.008$   & 0.85  \\ \hline 
6 & 10 & 3 & 12 & $0.435 \pm 0.009$   & 0.57  \\ \hline 
6 & 11 & 3 & 12 & $0.431 \pm 0.010$   & 0.29  \\ \hline 
 8 & 7 & 4 & 12 & $0.456 \pm 0.013$   & 0.37  \\ \hline 
 8 & 8 & 4 & 12 & $0.447 \pm 0.012$   & 0.54  \\ \hline 
 8 & 9 & 4 & 12 & $0.434 \pm 0.011$   & 0.53  \\ \hline 
8 & 10 & 3 & 12 & $0.433 \pm 0.008$   & 0.61  \\ \hline 
8 & 11 & 4 & 12 & $0.417 \pm 0.013$   & 0.63  \\ \hline 
\end{tabular}

\caption{Fitted scalar glueball mass in lattice units using the best
$t_{min}$ and $t_{max}$ for various choices
of $n$ and $s$ for the lattice 
$30 \times 32^2 \times 40$ at $\beta$ of $6.40$}
\label{tab:640z}
\end{center}
\end{table}
 
\begin{table}
\begin{center}
\begin{tabular}{ccccc}     \hline
$n$ & $s$ & mass & $\chi^2/d.o.f.$ &\\ \hline

4,5,6 & 1,2,3,4 & $0.978 \pm 0.013$    & 4.85 & \\ \hline
4,5,6 & 1,2,3   & $0.973 \pm 0.014$    & 3.70 & \\ \hline
4,5 & 1,2,3,4   & $0.975 \pm 0.014$    & 5.39 & \\ \hline
5,6 & 1,2,3,4   & $0.974 \pm 0.014$    & 6.46 & \\ \hline
4 & 1,2,3,4     & $0.965 \pm 0.014$    & 8.95  & \\ \hline
5 & 1,2,3,4     & $0.964 \pm 0.015$    & 9.90 & \\ \hline
6 & 1,2,3,4     & $0.966 \pm 0.015$    & 10.20 & \\ \hline
\hline
4,5,6 & 2,3     & $0.973 \pm 0.014$    & 5.07 & \\ \hline
\hline
5,6 & 2,3,4     & $0.973 \pm 0.014$    & 7.25 & \\ \hline
5 & 2,3,4       & $0.964 \pm 0.015$    & 14.80 & \\ \hline
\hline
4,5 & 2,3       & $0.966 \pm 0.014$    & 3.46  & \\ \hline
5,6 & 2,3       & $0.969 \pm 0.014$    & 6.04 & \\ \hline
\hline
4,5 & 2         & $0.957 \pm 0.015$    & 1.70 & \\ \hline
4,5 & 3         & $0.953 \pm 0.015$    & 0.20 & \\ \hline
5,6 & 3         & $0.953 \pm 0.015$    & 1.67 & \\ \hline
5,6 & 2         & $0.955 \pm 0.015$    & 0.35 & \\ \hline
4 & 2,3         & $0.955 \pm 0.014$    & 1.36  & \\ \hline
5 & 2,3         & $0.954 \pm 0.015$    & 0.22 & \\ \hline
6 & 2,3         & $0.955 \pm 0.015$    & 0.11 &  $\leftarrow$  \\ \hline

\end{tabular}

\caption{Scalar glueball mass in lattice units found by
combined fits to sets of $n$ and $s$ for the lattice $16^3 \times 24$ at
$\beta$ of $5.70$. The final set chosen is indicated by an arrow.}
\label{tab:570zf}
\end{center}
\end{table}

\begin{table}
\begin{center}
\begin{tabular}{ccccc}     \hline
$n$ & $s$ & mass & $\chi^2/d.o.f.$ \\ \hline

5,6,7 & 4,5,6 & $0.752 \pm 0.020$ & 2.35 & \\ \hline 
5,6,7 & 4,5 & $0.750 \pm 0.019$  & 1.75 & \\ \hline 
5,6,7 & 5,6 & $0.747 \pm 0.020$  & 0.98 & $\leftarrow$ \\ \hline 
5,6 & 4,5,6 & $0.752 \pm 0.019$  & 3.72 & \\ \hline 
6,7 & 4,5,6 & $0.751 \pm 0.020$  & 3.67 & \\ \hline 
5,6 & 4,5 & $0.749 \pm 0.019$  & 2.88 & \\ \hline 
5,6 & 5,6 & $0.745 \pm 0.020$  & 1.41 & \\ \hline 
6,7 & 4,5 & $0.747 \pm 0.020$  & 2.52 & \\ \hline 
6,7 & 5,6 & $0.746 \pm 0.020$  & 1.48 & \\ \hline 
6 & 5,6 & $0.742 \pm 0.020$  & 1.19 & \\ \hline 
7 & 5,6 & $0.748 \pm 0.020$  & 1.24 & \\ \hline 

\end{tabular}

\caption{Scalar glueball mass in lattice units found by
combined fits to sets of $n$ and $s$ for the lattice $12^3 \times 24$ at
$\beta$ of $5.93$. The final set chosen is indicated by an arrow.}
\label{tab:593szf}
\end{center}
\end{table}

\begin{table}
\begin{center}
\begin{tabular}{ccccc}     \hline
$n$ & $s$ & mass & $\chi^2/d.o.f.$ \\ \hline

5,6,7,8 & 4,5,6 & $0.781 \pm 0.011$  & 1.90 & $\leftarrow$ \\ \hline 
5,6,7,8 & 4,5 & $0.779 \pm 0.011$  & 2.77 & \\ \hline 
5,6,7,8 & 5,6 & $0.776 \pm 0.012$  & 1.99 & \\ \hline 
5,6,7 & 4,5,6 & $0.779 \pm 0.011$  & 2.14 & \\ \hline 
5,6,7 & 4,5 & $0.778 \pm 0.012$  & 3.24 & \\ \hline 
5,6,7 & 5,6 & $0.776 \pm 0.012$  & 2.29 & \\ \hline 
6,7,8 & 4,5,6 & $0.782 \pm 0.011$ & 2.32 & \\ \hline 
6,7,8 & 4,5 & $0.780 \pm 0.011$  & 3.50 & \\ \hline 
6,7,8 & 5,6 & $0.777 \pm 0.012$  & 2.00 & \\ \hline 
6,7 & 4,5,6 & $0.778 \pm 0.011$  & 3.01 & \\ \hline 
7,8 & 4,5,6 & $0.778 \pm 0.012$  & 1.68 & \\ \hline 
6 & 4,5,6 & $0.786 \pm 0.012$   & 0.21 & \\ \hline 
7 & 4,5,6 & $0.779 \pm 0.012$  & 0.32 & \\ \hline 
8 & 4,5,6 & $0.771 \pm 0.012$  & 0.86 & \\ \hline 

\end{tabular}

\caption{Scalar glueball mass in lattice units found by
combined fits to sets of $n$ and $s$ for the lattice $16^3 \times 24$ at
$\beta$ of $5.93$. The final set chosen is indicated by an arrow.}
\label{tab:593lzf}
\end{center}
\end{table}

\begin{table}
\begin{center}
\begin{tabular}{ccccc}     \hline
$n$ & $s$ & mass & $\chi^2/d.o.f.$ \\ \hline

7,8 & 7,8,9,10 & $0.559 \pm 0.017$  & 1.26 & $\leftarrow$ \\ \hline 
7,8 & 7,8,9 & $0.559 \pm 0.017$  & 1.70 & \\ \hline 
7,8 & 8,9,10 & $0.553 \pm 0.017$  & 1.46 & \\ \hline 
7 & 7,8,9,10 & $0.564 \pm 0.018$  & 0.58 & \\ \hline 
7 & 7,8,9 & $0.566 \pm 0.018$  & 0.60 & \\ \hline 
7 & 8,9,10 & $0.559 \pm 0.017$  & 0.40 & \\ \hline 
8 & 7,8,9,10 & $0.555 \pm 0.017$  & 0.63 & \\ \hline 
8 & 7,8,9 & $0.556 \pm 0.017$  & 0.91 & \\ \hline 
8 & 8,9,10 & $0.549 \pm 0.018$  & 0.30 & \\ \hline 
7,8 & 7,8 & $0.559 \pm 0.017$  & 2.79 & \\ \hline 
7 & 7,8 & $0.566 \pm 0.018$  & 1.18 & \\ \hline 
8 & 7,8 & $0.556 \pm 0.017$  & 1.80 & \\ \hline 
\end{tabular}

\caption{Scalar glueball mass in lattice units found by
combined fits to sets of $n$ and $s$ for the lattice $24^3 \times 36$ at
$\beta$ of $6.17$. The final set chosen is indicated by an arrow.}
\label{tab:617zf}
\end{center}
\end{table}

\begin{table}
\begin{center}
\begin{tabular}{ccccc}     \hline
$n$ & $s$ & mass & $\chi^2/d.o.f.$ \\ \hline

6,8 & 7,8,9,10,11 & $0.4416 \pm 0.0074$  & 3.72 & \\ \hline 
6,8 & 8,9,10,11 & $0.4416 \pm 0.0074$  & 4.13 & \\ \hline 
6,8 & 9,10,11 & $0.4413 \pm 0.0076$  & 5.37 & \\ \hline 
6 & 9,10,11 & $0.4477 \pm 0.0082$  & 9.90 & \\ \hline 
8 & 9,10,11 & $0.4363 \pm 0.0078$  & 2.55 & \\ \hline 
6,8 & 10,11 & $0.4321 \pm 0.0077$  & 1.14 & $\leftarrow$ \\ \hline 
6,8 & 10 & $0.4327 \pm 0.0076$  & 0.37 & \\ \hline 
6,8 & 11 & $0.4281 \pm 0.0096$  & 2.11 & \\ \hline 
6 & 10,11 & $0.4360 \pm 0.0087$  & 0.91 & \\ \hline 
8 & 10,11 & $0.4326 \pm 0.0077$  & 2.53 & \\ \hline 
\end{tabular}

\caption{Scalar glueball mass in lattice units found by
combined fits to sets of $n$ and $s$ for the lattice $30 \times 32^2 \times 40$ at
$\beta$ of $6.40$. The final set chosen is indicated by an arrow.}
\label{tab:640zf}
\end{center}
\end{table}

\begin{table}
\begin{center}
\begin{tabular}{ccc}     \hline
 $\beta$ &  lattice         & mass \\ \hline
 5.70    & $16^3 \times 24$ & $0.955 \pm 0.015$ \\ 
 5.93    & $12^3 \times 24$ & $0.747 \pm 0.020$ \\
         & $16^3 \times 24$ & $0.781 \pm 0.011$ \\
 6.17    & $24^3 \times 36$ & $0.559 \pm 0.017$ \\
 6.40    & $30 \times 32^2 \times 40$ 
                            & $0.4321 \pm 0.0077$ \\  \hline
\end{tabular}
\caption{Final scalar glueball mass values.}
\label{tab:finalz}
\end{center}
\end{table}

\begin{table}
\begin{center}
\begin{tabular}{cccccc}     \hline
$n$ & $s$ & $t_{min}$ & $t_{max}$ & mass & $\chi^2/d.o.f.$ \\ \hline

 5 & 4 & 2 & 5 & $1.260 \pm 0.036$   & 1.15  \\ \hline 
 5 & 5 & 2 & 5 & $1.226 \pm 0.029$   & 0.70  \\ \hline 
 5 & 6 & 2 & 5 & $1.224 \pm 0.031$   & 0.41  \\ \hline 
 6 & 4 & 2 & 5 & $1.250 \pm 0.034$   & 1.00  \\ \hline 
 6 & 5 & 2 & 5 & $1.218 \pm 0.029$   & 0.66  \\ \hline 
 6 & 6 & 2 & 5 & $1.213 \pm 0.029$   & 0.42  \\ \hline 
 7 & 4 & 2 & 5 & $1.245 \pm 0.032$   & 0.59  \\ \hline 
 7 & 5 & 2 & 5 & $1.209 \pm 0.029$   & 0.45  \\ \hline 
 7 & 6 & 2 & 5 & $1.206 \pm 0.028$   & 0.31  \\ \hline 
\end{tabular}
\caption{Fitted tensor glueball mass in lattice units using the best
$t_{min}$ and $t_{max}$ for various choices of $n$ and $s$ for the
lattice $12^3 \times 24$ at $\beta = 5.93$.}
\label{tab:593st}
\end{center}
\end{table}

\begin{table}
\begin{center}
\begin{tabular}{cccccc}     \hline
$n$ & $s$ & $t_{min}$ & $t_{max}$ & mass & $\chi^2/d.o.f.$ \\ \hline

 5 & 4 & 1 & 5 & $1.327 \pm 0.013$   & 0.51 \\ \hline 
 5 & 4 & 2 & 5 & $1.284 \pm 0.043$   & 0.34 \\ \hline 
 5 & 5 & 1 & 5 & $1.285 \pm 0.011$   & 0.72 \\ \hline 
 5 & 5 & 2 & 5 & $1.270 \pm 0.039$   & 1.02 \\ \hline 
 5 & 6 & 1 & 5 & $1.284 \pm 0.011$   & 1.54 \\ \hline 
 5 & 6 & 2 & 5 & $1.252 \pm 0.042$   & 2.04 \\ \hline 
 6 & 4 & 1 & 5 & $1.302 \pm 0.012$   & 0.53 \\ \hline 
 6 & 4 & 2 & 5 & $1.282 \pm 0.043$   & 0.70 \\ \hline 
 6 & 5 & 1 & 5 & $1.267 \pm 0.011$   & 0.83 \\ \hline 
 6 & 5 & 2 & 5 & $1.264 \pm 0.037$   & 1.24 \\ \hline 
 6 & 6 & 1 & 5 & $1.264 \pm 0.011$   & 1.54 \\ \hline 
 6 & 6 & 2 & 5 & $1.243 \pm 0.037$   & 2.18 \\ \hline 
 7 & 4 & 1 & 5 & $1.282 \pm 0.012$   & 0.79 \\ \hline 
 7 & 4 & 2 & 5 & $1.278 \pm 0.042$   & 1.19 \\ \hline 
 7 & 5 & 1 & 5 & $1.252 \pm 0.011$   & 0.95 \\ \hline 
 7 & 5 & 2 & 5 & $1.255 \pm 0.036$   & 1.42 \\ \hline 
 7 & 6 & 1 & 5 & $1.249 \pm 0.011$   & 1.64 \\ \hline 
 7 & 6 & 2 & 5 & $1.233 \pm 0.035$   & 2.37 \\ \hline 
\end{tabular}
\caption{Fitted tensor glueball mass in lattice units for various choices
of $n$, $s$, $t_{min}$ and $t_{max}$ for the lattice $16^3 \times 24$ at $\beta = 5.93$.}
\label{tab:593lt}
\end{center}
\end{table}

\begin{table}
\begin{center}
\begin{tabular}{cccccc}     \hline
$n$ & $s$ & $t_{min}$ & $t_{max}$ & mass & $\chi^2/d.o.f.$ \\ \hline

 7 & 7 & 3 & 7 & $0.861 \pm 0.029$    & 0.75  \\ \hline 
 7 & 8 & 3 & 7 & $0.830 \pm 0.027$    & 0.24  \\ \hline 
 7 & 9 & 3 & 7 & $0.820 \pm 0.028$    & 0.07  \\ \hline 
7 & 10 & 3 & 7 & $0.813 \pm 0.034$   & 0.53  \\ \hline 
 8 & 7 & 2 & 7 & $0.870 \pm 0.010$    & 0.46  \\ \hline 
 8 & 8 & 3 & 7 & $0.819 \pm 0.026$    & 0.49  \\ \hline 
 8 & 9 & 2 & 7 & $0.839 \pm 0.009$    & 0.27  \\ \hline 
8 & 10 & 3 & 7 & $0.815 \pm 0.031$   & 0.08  \\ \hline 
\end{tabular}
\caption{Fitted tensor glueball mass in lattice units using the best
$t_{min}$ and $t_{max}$ for various choices of $n$ and $s$ for the
lattice $24^3 \times 36$ at $\beta = 6.17$.}
\label{tab:617t}
\end{center}
\end{table}

\begin{table}
\begin{center}
\begin{tabular}{cccccc}     \hline
$n$ & $s$ & $t_{min}$ & $t_{max}$ & mass & $\chi^2/d.o.f.$ \\ \hline

 6 & 6 & 4 & 9  & $0.680 \pm 0.043$  & 0.71  \\ \hline 
 6 & 6 & 5 & 9  & $0.642 \pm 0.073$  & 0.84  \\ \hline 
 6 & 7 & 4 & 9  & $0.660 \pm 0.034$  & 0.65  \\ \hline 
 6 & 7 & 5 & 9  & $0.643 \pm 0.063$  & 0.84  \\ \hline 
 6 & 8 & 4 & 9  & $0.652 \pm 0.029$  & 0.33  \\ \hline 
 6 & 8 & 5 & 9  & $0.638 \pm 0.059$  & 0.41  \\ \hline 
 6 & 9 & 3 & 9  & $0.688 \pm 0.016$  & 0.58  \\ \hline 
 6 & 9 & 4 & 9  & $0.657 \pm 0.030$  & 0.38  \\ \hline 
 6 & 10 & 3 & 9 & $0.675 \pm 0.016$  & 0.56  \\ \hline 
 6 & 10 & 4 & 9 & $0.670 \pm 0.032$  & 0.69  \\ \hline 
 6 & 11 & 3 & 9 & $0.660 \pm 0.017$  & 0.31  \\ \hline 
 6 & 11 & 4 & 9 & $0.665 \pm 0.033$  & 0.38  \\ \hline 
 8 & 6 & 4 & 9  & $0.658 \pm 0.038$  & 0.46  \\ \hline 
 8 & 6 & 5 & 9  & $0.651 \pm 0.068$  & 0.60  \\ \hline 
 8 & 7 & 4 & 9  & $0.636 \pm 0.030$  & 0.50  \\ \hline 
 8 & 8 & 4 & 9  & $0.634 \pm 0.027$  & 0.19  \\ \hline 
 8 & 9 & 4 & 9  & $0.637 \pm 0.026$  & 0.25  \\ \hline 
 8 & 10 & 3 & 9 & $0.660 \pm 0.014$  & 0.35  \\ \hline 
 8 & 10 & 4 & 9 & $0.653 \pm 0.027$  & 0.42  \\ \hline 
 8 & 11 & 3 & 9 & $0.646 \pm 0.015$  & 0.15  \\ \hline 
 8 & 11 & 4 & 9 & $0.653 \pm 0.026$  & 0.17  \\ \hline 
\end{tabular}
\caption{Fitted tensor glueball mass in lattice units 
for various choices of $n$, $s$, $t_{min}$ and $t_{max}$ for the
lattice $32^2 \times 30 \times 40$ at $\beta = 6.4$}
\label{tab:640t}
\end{center}
\end{table}

\begin{table}
\begin{center}
\begin{tabular}{ccccc}     \hline
$n$ & $s$ &  mass & $\chi^2/d.o.f.$ &\\ \hline
5,6,7 & 4,5,6  & $1.188 \pm 0.025$ & 2.13 & $\leftarrow$ \\ \hline 
5,6,7 & 4,5  & $1.207 \pm 0.028$ & 1.73  & \\ \hline 
5,6,7 & 5,6  & $1.195 \pm 0.025$ & 1.33  & \\ \hline 
5,6 & 4,5   & $1.206 \pm 0.029$ & 1.97   & \\ \hline 
5,6 & 5,6  & $1.199 \pm 0.026$ & 1.29   & \\ \hline 
6,7 & 4,5  & $1.205 \pm 0.029$ & 1.90   & \\ \hline 
6,7 & 5,6  & $1.207 \pm 0.027$ & 0.88   & \\ \hline 
\end{tabular}
\caption{Tensor glueball mass in lattice units found by
combined fits to sets of $n$ and $s$ for the lattice $12^3 \times 24$ at
$\beta$ of $5.93$. The final set chosen is indicated by an arrow.}
\label{tab:593stf}
\end{center}
\end{table}

\begin{table}
\begin{center}
\begin{tabular}{ccccccc}     \hline
$n$ & $s$ & $t_{min}$ & $t_{max}$ & mass & $\chi^2/d.o.f.$ &\\ \hline

5,6,7 & 4,5 &  1 & 5 &  $1.256 \pm 0.011$  & 31.87 & \\ \hline 
5,6 & 4,5 &    1 & 5 &  $1.260 \pm 0.011$  & 44.45 & \\ \hline \hline
5,6,7 & 4,5,6 & 2 & 5 & $1.234 \pm 0.034$  & 0.60  & $\leftarrow$ \\ \hline 
5,6,7 & 4,5 &   2 & 5 & $1.252 \pm 0.036$  & 0.48  & \\ \hline 
5,6,7 & 5,6 &   2 & 5 &  $1.236 \pm 0.036$  & 0.86  & \\ \hline 
5,6 & 4,5 &     2 & 5 &  $1.254 \pm 0.036$  & 0.59  & \\ \hline 
5,6 & 5,6 &     2 & 5 &  $1.239 \pm 0.035$  & 1.01  & \\ \hline 
6,7 & 4,5 &     2 & 5 &  $1.256 \pm 0.037$  & 0.60  & \\ \hline 
6,7 & 5,6 &     2 & 5 &  $1.239 \pm 0.034$  & 1.13  & \\ \hline 
\end{tabular}
\caption{Tensor glueball mass in lattice units found by
combined fits to sets of $n$ and $s$ for the lattice $16^3 \times 24$ at
$\beta$ of $5.93$. The final set chosen is indicated by an arrow.}
\label{tab:593ltf}
\end{center}
\end{table}

\begin{table}
\begin{center}
\begin{tabular}{ccccc}     \hline
$n$ & $s$ & mass & $\chi^2/d.o.f.$ &\\ \hline

7,8 & 7,8,9,10 & $0.856 \pm 0.010$ & 2.84 & \\ \hline 
7,8 & 7,8,9   & $0.856 \pm 0.010$ & 3.83  & \\ \hline 
7,8 & 8,9,10  & $0.838 \pm 0.012$ & 1.24  & $\leftarrow$ \\ \hline 
7,8 & 7,8    & $0.865 \pm 0.011$ & 4.19   & \\ \hline 
7,8 & 8,9    & $0.838 \pm 0.012$ & 1.72   & \\ \hline 
7,8 & 9,10   & $0.840 \pm 0.012$ & 0.21   & \\ \hline 
\end{tabular}
\caption{Tensor glueball mass in lattice units found by
combined fits to sets of $n$ and $s$ for the lattice $24^3 \times 36$ at
$\beta = 6.17$ The final set chosen is indicated by an arrow.}
\label{tab:617tf}
\end{center}
\end{table}

\begin{table}
\begin{center}
\begin{tabular}{ccccccc}     \hline
$n$ & $s$ & mass & $t_{min}$ & $t_{max}$ & $\chi^2/d.o.f.$ &\\ \hline

6,8 & 6,7,8,9,10,11 & 3,4 & 9 & $0.655 \pm 0.012$ & 1.81  & \\ \hline 
6,8 & 6,7,8,9,10   & 3,4 & 9  & $0.659 \pm 0.014$ & 2.04  & \\ \hline 
6,8 & 7,8,9,10,11  & 3,4 & 9  & $0.655 \pm 0.012$ & 2.06  & \\ \hline 
6,8 & 6,7,8,9     & 3,4 & 9   & $0.676 \pm 0.015$ & 1.52  & \\ \hline 
6,8 & 7,8,9,10    & 3,4 & 9   & $0.659 \pm 0.013$ & 2.41  & \\ \hline 
6,8 & 8,9,10,11   & 3,4 & 9   & $0.655 \pm 0.012$ & 2.22  & \\ \hline 
\hline
6,8 & 6,7,8,9,10,11 & 4,5 & 9 & $0.631 \pm 0.022$ & 0.56  & $\leftarrow$ \\ \hline 
6,8 & 6,7,8,9,10   & 4,5 & 9  & $0.635 \pm 0.024$ & 0.65  & \\ \hline 
6,8 & 7,8,9,10,11  & 4,5 & 9  & $0.632 \pm 0.021$ & 0.60  & \\ \hline 
6,8 & 6,7,8,9     & 4,5 & 9   & $0.627 \pm 0.024$ & 0.56  & \\ \hline 
6,8 & 7,8,9,10    & 4,5 & 9   & $0.635 \pm 0.025$ & 0.73  & \\ \hline 
6,8 & 8,9,10,11   & 4,5 & 9   & $0.630 \pm 0.022$ & 0.65  & \\ \hline 
\end{tabular}
\caption{Tensor glueball mass in lattice units found by
combined fits to sets of $n$ and $s$ for the lattice $32^2 \times 30
\times 40$ at $\beta = 6.4$ The final set chosen is indicated by an
arrow.}
\label{tab:640tf}
\end{center}
\end{table}

\begin{table}
\begin{center}
\begin{tabular}{ccc}     \hline
 $\beta$ &  lattice         & mass \\ \hline
 5.93    & $12^3 \times 24$ & $1.188 \pm 0.025$ \\
         & $16^3 \times 24$ & $1.234 \pm 0.034$ \\
 6.17    & $24^3 \times 36$ & $0.838 \pm 0.012$ \\
 6.40    & $30 \times 32^2 \times 40$ 
                            & $0.631 \pm 0.022$ \\  \hline
\end{tabular}
\caption{Final tensor glueball mass values.}
\label{tab:finalt}
\end{center}
\end{table}

\begin{table}
\begin{center}
\begin{tabular}{ccc}     \hline
 $\beta$ &  $\alpha_{\overline{MS}}$  &  $\Lambda^{(0)}_{\overline{MS}} a$ \\ \hline
5.700 &   0.14557 &  0.16612 \\ 
5.930 &   0.13180 &  0.11444 \\ 
6.170 &   0.12183 &  0.08265 \\ 
6.400 &   0.11407 &  0.06177 \\ \hline
\end{tabular}
\caption{For each $\beta$, mean-field improved $\alpha_{\overline{MS}}$
and $\Lambda^{(0)}_{\overline{MS}} a$ obtained from the 2-loop
Callan-Synamzik equation.}
\label{tab:lambdas}
\end{center}
\end{table}

\begin{table}
\begin{center}
\begin{tabular}{cc}     \hline
$m_0/\Lambda^{(0)}_{\overline{MS}}$ & $7.016 \pm 0.167$ \\
$m_2/\Lambda^{(0)}_{\overline{MS}}$ & $9.65 \pm 0.36$ \\
$m_2/m_0$ & $1.375 \pm 0.066$ \\
$\Lambda^{(0)}_{\overline{MS}}/m_{\rho}$ & $0.305 \pm 0.008$ \\
$m_{\rho}$ & 770 MeV \\
$\Lambda^{(0)}_{\overline{MS}}$ & $234.9 \pm 6.2$ MeV \\
$m_0$ & $1648 \pm 58$ MeV \\
$m_2$ & $2267 \pm 104$ MeV \\ \hline
\end{tabular}
\caption{Continuum limit scalar and tensor glueball masses 
and their conversion to MeV.}
\label{tab:contlim}
\end{center}
\end{table}

\begin{thebibliography}{9}
\bibitem{masses} H.\ Chen, J.\ Sexton, A.\ Vaccarino
and D.\ Weingarten, Nucl.\ Phys.\ B (Proc.\ Suppl.) {\bf 34}, 357 (1994).
\bibitem{review} D.\ Weingarten, Nucl.\ Phys.\ B (Proc.\ Suppl.) 
{\bf 34}, 29 (1994).
\bibitem{glbdecay} J.\ Sexton, A.\ Vaccarino and D.\ Weingarten, 
Phys.\ Rev.\ Letts.\ 75 (1995) 4563; Nucl.\ Phys.\ B (Proc.\ Suppl.)
{\bf 47},
128 (1996).
\bibitem{contadv} Don Weingarten, in Continuous Advances in QCD
1996,(edited by M.\ Polikarpov), World Scientific, Singapore, 1996.
\bibitem{lat98mix} W.\ Lee and D.\ Weingarten, Nucl.\ Phys.\ B (Proc.\
Suppl.) {\bf 73}, 249 (1999).
\bibitem{valence} D.\ H.\ Weingarten, Phys.\ Lett.\ 109B (1982) 57;
Nucl.\ Phys.\ {\bf B215 [FS7]}, 1 (1983).
\bibitem{hadrons} F.\ Butler, H.\ Chen, J.\ Sexton, A.\ Vaccarino and
D.\ Weingarten, Phys.\ Rev.\ Letts.\ {\bf 70}, 2849 (1993);
Nucl.\ Phys.\ {\bf B430} (1994) 179; Nucl.\ Phys.\ {\bf B421},
217 (1994).
\bibitem{tsukuba} K.\ Kanaya, {\it et al.}, Nucl.\ Phys.\ B (Proc.\ Suppl.)
{\bf 63A-C}, 161 (1998).
\bibitem{lat96} D.\ Weingarten, Nucl.\ Phys.\ B (Proc.\ Suppl.) {\bf
53}, 232 (1997).
\bibitem{livertal} G.\ Bali, K.\ Schilling,
A.\ Hulsebos, A.\ Irving, C.\ Michael, P.\ Stephenson, 
Phys.\ Lett.\ {\bf B 309}, 378 (1993).
\bibitem{Morningstar} C.\ Morningstar and M.\ Peardon, Phys.\ Rev.\ D
{\bf 56}, 4043 (1997); hep-lat/9901004, to appear in Phys.\ Rev.\ D.
\bibitem{Gupta} R.\ Gupta {\it et al.}, Phys.\ Rev.\ D {\bf 43}, 2301 (1993).
\bibitem{APE}
M.\ Falcioni {\it et al.}, Nucl.\ Phys.\ {\bf B251}, 624 (1985);
M.\ Albanese {\it et al.}, Phys.\ Lett.\ {\bf B192}, 163 (1985).
\bibitem{Tepper} M.\ Teper, Phys.\ Lett.\ {\bf B 183}, 345 (1987).
\bibitem{Brown} F.\ R.\ Brown and T.\ J.\ Woch, Phys.\ Rev.\ Lett.\ {\bf 58},
2394 (1987).
\bibitem{Efron} B.\ Efron, The Jackknife, the Bootstrap and Other
Resampling Plans, Society for Industrial and Applied Mathematics,
Philadelphia, 1982.
\bibitem{vanBaal} P.\ van Baal and A.\ Kronfeld, Nucl.\ Phys.\ B (Proc.\
Suppl.) {\bf 9}, 227 (1989).
\bibitem{Luscher}
M.\ L\"{u}scher, Comm.\ Math.\ Phys.\ {\bf 104}, 177 (1986).
\bibitem{Schierholz}
G.\ Schierholz, Nucl.\ Phys.\ B (Proc. Suppl.) {\bf 9}, 244 (1989).
\bibitem{Lepage} G.\ P.\ Lepage and P.\ Mackenzie, Phys.\ Rev.\ D {\bf 48},
2250 (1993).
\bibitem{Gottlieb} S.\ Gottlieb, private communication.
\bibitem{Sommer} R.\ Sommer, Nucl.\ Phys.\ {\bf B411}, 839 (1994).
\end{thebibliography}
\end{document}